\documentclass[12 pt]{article}
\usepackage{lscape}
\usepackage{pdfpages}

\usepackage[utf8]{inputenc}

\usepackage{amssymb,amsfonts,amsmath,mathrsfs}
\usepackage[nohead]{geometry}
\usepackage[singlespacing]{setspace}
\usepackage[bottom]{footmisc}
\usepackage{indentfirst}
\usepackage{endnotes}
\usepackage{graphicx}
\usepackage{adjustbox}
\usepackage{chngcntr}
\usepackage{bm}

\usepackage{titlesec}
\usepackage{etoolbox}

\makeatletter
\def\@biblabel#1{\hspace*{-\labelsep}}

\makeatother
\usepackage[round]{natbib}
\usepackage{xcolor}
\usepackage[
  bookmarks=true,
  bookmarksopen=true,
  breaklinks=true,
  colorlinks=true,
  citecolor=tred,
  linkcolor=tred,
  urlcolor=tred,
]{hyperref}

\usepackage{booktabs}

\usepackage{multirow}

\numberwithin{equation}{section}

\usepackage{titlesec}
\titlespacing*{\section}{0pt}{0.09 in}{0.09 in}
\titlespacing*{\subsection}{0pt}{0.07 in}{0.07 in}

\usepackage{etoolbox}
\newcommand{\zerodisplayskips}{%
  \setlength{\abovedisplayskip}{0.05 in}
  \setlength{\belowdisplayskip}{0.05 in}
  \setlength{\abovedisplayshortskip}{0.05 in}
  \setlength{\belowdisplayshortskip}{0.05 in}}
\appto{\normalsize}{\zerodisplayskips}
\appto{\small}{\zerodisplayskips}
\appto{\footnotesize}{\zerodisplayskips}

\allowdisplaybreaks

\usepackage{tikz}
\usetikzlibrary{calc}
\usetikzlibrary{decorations.pathreplacing,angles,quotes} 
\usepackage{pgfplots}
\pgfplotsset{compat=1.8}
\usepackage{mathtools}

\renewcommand{\[}{\left[}

\definecolor{tred}{RGB}{200,50,80}
\definecolor{tgreen}{RGB}{80,180,80}
\definecolor{tblue}{RGB}{50,50,250}
\definecolor{tpupple}{RGB}{150,50,200}
\usepackage{subcaption}

\usepackage{float}
\usepackage{kotex}
\usepackage{threeparttable} 
\usepackage{caption}
\usepackage{tikz}
\usepackage{verbatim}
\usepackage{xcolor}

\usepackage{csquotes}

\begin{document}

\begin{singlespace}

\title{Implicit Bias against a Capitalistic Society Predicts Market Earnings\footnote{We would like to thank seminar participants at the AEI-Five Joint Conference and the World Congress of Comparative Economics 2017, the Asian Conference on Applied Micro and Econometrics 2017, University of Cambridge, Hanyang University, Seoul National University, and Yonsei University for valuable comments. J. Lee's work was supported by the Ministry of Education of the Republic of Korea, the National Research Foundation of Korea (NRF-2020S1A3A2A02104190) and Overseas Training Expenses for Humanities and Social Sciences through Seoul National University.}}

\author{Syngjoo Choi \\ SNU \and Kyu Sup Hahn \\ SNU \and Byung-Yeon Kim \\ SNU \and Eungik Lee \\ NYU \and Jungmin Lee \\ SNU \& IZA \and Sokbae Lee \\ Columbia \& IFS}

\date{February 21, 2023 \\
\vspace{0.1 in}
}

\maketitle

\sloppy

\onehalfspacing

\begin{abstract}
\noindent 
This paper investigates whether ideological indoctrination by living in a communist regime relates to low economic performance in a market economy. We recruit North Korean refugees and measure their implicit bias against South Korea by using the Implicit Association Test. Conducting double auction and bilateral bargaining market experiments, we find that North Korean refugees with a larger bias against the capitalistic society have lower expectations about their earning potential, exhibit trading behavior with lower target profits, and earn less profits. These associations are robust to conditioning on correlates of preferences, human capital, and assimilation experiences. \\

\end{abstract}
\end{singlespace}

\strut


\onehalfspacing 



\newpage

\section{Introduction}
Societies are founded on a particular set of social values. Citizens are inculcated with these social values through various channels, such as education from parents and teachers, mass media, and legal institutions (\citealp{lott:1999}; \citealp{bowlesgintis:2011}; \citealp{cantoniyuchtman:2013}). A famous example is the Soviet Union, which promoted propaganda about creating \textit{Homo sovieticus}---a new kind of human being who is selfless, learned, healthy, muscular, and enthusiastic in spreading the Socialist Revolution (\citealp{zinoviev:1983}; \citealp{shilleretal:1991}). This large-scale attempt of so-called human remodelling impacts people's beliefs and attitudes. Even in decades after the collapse of the Soviet regime, we can still see the legacies of ideological indoctrination (\citealp{alesinaandfuchs:2007}; \citealp{popelchesandtucker:2017}). Diverging political attitudes exist toward \textit{laissez faire} and state interventions across countries depending on their historical experiences (\citealp{benabou:2008}; \citealp{cantonietal:2017}).

In this study, we examine whether ideological indoctrination by living in an authoritarian communist regime relates to low economic performance in a market economy. To answer this question, we focus on a unique group of people---North Korean (NK) refugees residing in South Korea (SK).\footnote{We use the acronyms of NK and SK to refer to both a person (North Korean and South Korean) and the country (North Korea and South Korea).} Since Korea was divided into North and South in 1948 as a consequence of the Korean War, two contrasting regimes have emerged and moved forward in the two parts of the Korean peninsula; a market economy and democracy in SK versus a centrally panned economy and authoritarian communist regime in NK. NK leaders decided to insulate their country from the rest of the world and inculcated their citizens with their own political ideology. One of the key components in NK's political propaganda is the implantation of antipathy toward capitalism. The \textit{Juche} ideology, translated into “we-centeredness,” acclaims the Great Fight against capitalist imperialism and identifies the U.S. and SK as primary enemies. As a consequence, NKs are gradually instilled with the creed that ``socialism is superior to capitalism" and that ``South Korea and the U.S. are the biggest enemies." Having this deeply embedded mindset, such indoctrination could have a persisting effect on their economic behavior, even long after leaving their home country and resettling in SK---more than 11 years in our sample.\footnote{We provide a detailed explanation of indoctrination in NK in Online Appendix \textcolor{blue}{D}.} 


There are two major methodological challenges in this study. The first  concerns a measurement problem---how to measure NK refugees' bias against SK society. It is not only difficult to define but also deeply rooted in the mindset, and the deeper it is located, the more difficult it is to extract. We adapt the Implicit Association Test (IAT) to our context, a test that is widely used for implicit attitudes in social and behavioral sciences (see, e.g., \citealp{greenwald2009understanding}; \citealp{beaman2009powerful}; \citealp{lowes2015understanding}; \citealp{jost2009existence}; \citealp{stanley2011implicit}). 

Using the IAT, we measure the antipathy toward SK (or the affinity for NK) and interpret the test's score as a proxy for implicit bias against the capitalist society of SK. To stimulate the implicit bias, our design presents four national symbols of each country: a country map, the national flag, a photo image of a soldier in the national army uniform, and the official name of the country. We ask subjects to associate the symbols with words of good or bad attributes, while a timer unobserved by subjects is running. This allows us to measure the strength of an individual's implicit bias against a country. For example, NK refugees who are more biased against SK would take longer to respond correctly to a sorting problem when SK is associated with a word with a positive connotation. 

The second challenge is an identification problem. Our objective is to examine the association between  bias against the capitalistic society and market earnings. Studies with observational data are subject to identification; this is because individuals' market outcomes, such as earnings and wealth, are determined by numerous other factors, such as preferences, skills, and constraints. For example, NK refugees who are more biased against SK may also be more disadvantaged in the SK labor market with other reasons. 

To resolve this identification problem, we resort to a controlled laboratory market where participants seek profits on a level playing field without facing any differential market constraints. Another advantage of the market experiment is that we could control for the participants' economic preferences by assigning buyers’ values and sellers’ costs randomly. The market experiment utilizes continuous double auction institutions {\`a} la \citet{smith:1962}. We design two types of markets with this institution---double auction (DA) markets with multiple buyers and sellers and bilateral bargaining (BB) markets with only one buyer and one seller. We use the amount of money earned in the market experiment as the outcome of main interest. In addition, to mitigate the concerns of omitted variable bias, we collect various potential confounding factors, including a set of human capital measures, economic preferences, and information on NK refugees' life experiences back in NK and settlement experiences in SK. 



We find that NK refugees with a larger bias against SK earn less profits in both DA and BB markets. This finding holds robust to conditioning on the potential confounding factors. To understand mechanisms underlying that association, using the Becker-DeGroot-Marschak (BDM) design, we measure subjects' expectations about the profits they can earn in the market and show that NK refugees with a larger bias have lower expectations about their earning potential. Furthermore, using micro-level data of trading behavior, we show that NK refugees with a larger bias tend to engage in trading with a lower target profit, which in turn results in lower actual profits. These findings lend strong support to the hypothesis that NK refugees with a larger bias against the capitalistic society underperform in the market due to their trading behavior with low aspiration. 


Our study contributes to the literature on the long-term effect of life experiences, especially the experience of growing up in a communist regime, on later-life attitudes, beliefs, and economic behaviors (e.g., \citealp{shilleretal:1991};  \citealp{alesinaandfuchs:2007};  \citealp{cantonietal:2017}; \citealp{popelchesandtucker:2017}; \citealp{laudenbach2019emotional}). Particularly, \citet{fuchsandmasella:2016} and \citet{laudenbachetal:2019} compared East and West Germans after their reunification and found that longer exposure to communism is associated with lower investment in human capital, lower wages in the labor market, and poor financial decision. However, as \citet{beckeretal:2020} pointed out, East and West Germans are not comparable in various aspects; thus these studies with observation data face the issue of identification. 


Although we also utilize people from a communist country, our study does not suffer the same concern, given that we focus on heterogeneity among NK refugees rather than comparing them with SK natives and use a controlled laboratory outcome. In this sense, in terms of research design, our study is in line with \citet{laudenbach2019emotional}, who looked at  variations among East Germans in their emotional tagging to the communist system and whether positive or negative emotional tags shape long-run beliefs about capitalism versus communism. Our study provides novel insight into this literature by showing that NK refugees' bias against the capitalistic society is closely related to their market behaviors and economic performance. 







The rest of this paper proceeds as follows. Section~\ref{sec3} explains our experiments and surveys, including the IAT method. Section~\ref{sec4} presents the main findings from our market experiments. In Section~\ref{sec5}, we explore mechanisms underlying our findings. Section~\ref{sec6} contains a discussion about our results.

\section{Research Design and Data Collection}
\label{sec3}

\subsection{Recruiting Subjects}

We recruited subjects in collaboration with Nielsen Korea between late June and mid-July 2016. To construct a representative sample of NK refugees residing in SK, we used the stratified sampling method of recruiting NK refugees in terms of age, gender, and year of entry into SK. Then, as a benchmark group, we recruited SK natives to match the distribution of age and gender with those of NK refugees. 

In total, 289 NK refugees and 295 SK natives participated in our study. They were assigned to 20 sessions; four sessions with only SK native participants, three sessions with only NK refugees, and 13 sessions with both groups of subjects together. Forming the groups of eight subjects for the DA market experiment, we have 17 groups with only NK refugees (NK-only Market), 17 groups with only SK natives participants (SK-only Market), and 41 groups with both NK refugees and SK natives (Mixed Market).\footnote{Online Appendix \textcolor{blue}{C.2} reports a balance test results of NK refugees between NK only and Mixed markets. Overall NK refugees are well-balanced in many aspects despite small samples.} In most of mixed markets (36 out of 41), we have the same number of NK refugees and South Koreans and assigned two from each of NK refugees and South Koreans to play the role of buyers. The final sample consists of 287 NK refugees and 287 SK natives by eliminating 10 subjects, 6 replacement subjects from Nielsen Korea to form groups and 4 subjects who exhibited too erratic behavior in the IAT (see \citealp{noseketal:2014}). On average, subjects earned 68,000 Korean Won (KRW), including 40,000 KRW as a participation fee (as of July 2016, the exchange rate was approximately 1,000 KRW = 1 USD). 
We had a relatively high participation fee because most of the participants are non-students; thus, we need to provide a larger incentive compared to that given typically to a sample of university students. The experiment and survey lasted approximately 2 hours.

 Online Appendix \textcolor{blue}{C.1} shows the basic sociodemographic characteristics of our NK and SK subjects. In anticipation of exploiting the variation within NK subjects, it is worth noting that NK refugees are heterogeneous in their length of stay in NK and SK. We expect that the longer NK refugees have stayed in NK, the stronger bias they exhibit against SK. On average, they spent 27 years in NK and approximately 12 years outside of the country. The average duration of stay in SK is 7.3 years. The variation in the length of stay in NK is large: the minimum number of years in NK is 8 and the maximum is 52.   



\subsection{Market Experiments}

We conducted two market experiments, starting with the standard continuous DA market. At the beginning of the experiment, the computer randomly forms groups of eight anonymous subjects. These groups participate in seven rounds of the DA market. We run two practice rounds before the first round to ensure that all subjects understand the experiment rules. After the DA market, the subjects participate in three rounds of a one-on-one BB market with the same format as that of the DA market, by being randomly matched with another subject in the same group. All the experiments are computerized and conducted using the experimental software z-Tree (\citealp{fischbacherandurs:2007}). Instructions regarding all the experiments and surveys are reported in Online Appendix \textcolor{blue}{A}.

\medskip
\textbf{Double auction market.} We use the standard market experiment tool with continuous DA institution (e.g., \citealp{smith:1962}). Each market consists of four buyers and four sellers. The subjects are randomly assigned to the roles of buyers and sellers, which are fixed throughout seven rounds of the market. Each seller has one unit of a fictitious commodity with a reservation price in KRW, known only to herself, which represents the minimum price at which she is willing to relinquish one unit of the commodity. Each buyer receives a value in KRW, known only to herself, which represents the maximum willingness to pay for one unit of the fictitious commodity. Four buyers in a market have four different values---13,000, 17,000, 21,000, and 25,000---and four sellers have four different reservation prices---2,000, 6,000, 10,000, and 14,000. Although the role of a buyer or a seller is fixed throughout seven rounds of the market, each buyer’s value or each seller’s reservation price is randomly drawn from the sets in each round.  The notion of competitive equilibrium predicts three units of the commodity to be traded at a price range between 13,000 and 14,000. 

Each market lasts for 180 seconds. Both buyers and sellers are allowed to freely post the prices they want to buy or sell within each round. We note that they can update their posted prices as many times as they want during the given time. These posted prices are public information within a group. If the subjects want to make a transaction on a posted price, they can simply click that posted price and then press the OK button. If a buyer and a seller make a transaction, then they are out of the market, since only one transaction is possible for each subject. The payoff of a buyer is the difference between her value and the transaction price. The payoff of a seller is the difference between the transaction price and her reservation price. We call both payoffs as profits. If they do not make a transaction, then they have zero profit in the market. 

\medskip
\textbf{Eliciting expectation about profits.} Before starting the last round of the DA market, we elicit subjects' (ex-ante) expected value of participating in the last-round market, by using the Becker-DeGroot-Marschak (BDM) design (\citealp{beckeretal:1964}). Specifically, the subjects are presented with ten-row decision problems and asked to choose whether their earnings at the last round are determined by their performance in that market or given a fixed amount of money. As the row goes down, the fixed amount of money increases from 1,000 KRW to 10,000 KRW in increments of 1,000 KRW. We enforce a single crossing point to measure their willingness to participate in the market. After this elicitation procedure, they play the last round of the market. When the market ends, the computer randomly selects one out of the ten rows. The payoff obtained by a subject at the last round is determined by her choice of that selected row between a fixed amount of money and the profit earned in the market.

At the end of the session, each subject's earnings from the DA market were determined by the sum of payoffs obtained in two randomly drawn rounds among these seven rounds. 

\medskip
\textbf{Bilateral bargaining market.} The process of the BB market is identical to the DA market except that there are only one buyer and one seller. We choose to conduct this design to obtain a measure of individuals' performance in the simplest situation without spillover beyond bilateral interaction and minimize loads of cognition in processing information.

In each group of four buyers and four sellers, four pairs of one buyer and one seller are randomly matched in each round to play three rounds of two-person BB market. 
In each round of BB, the buyer and seller randomly receive one value that was used in the DA market experiment. 

The amount of surplus to be negotiated by the pair is equal to the difference between the value assigned to the buyer and the reservation price assigned to the seller. They negotiate for 90 seconds over the unknown surplus by submitting the prices at which they want to transact. If the pair would agree within the time limit, the buyer would obtain earnings equal to the difference between her value and the price on which the pair agreed, and the seller's earnings would be equal to the difference between the price on which the pair agreed and her reservation price. At the end of the session, one of the three rounds of BB is randomly chosen for the actual payment in this part of the experiment.

\subsection{Implicit Association Test}


At the end of the session, we conduct the IAT to measure each subject's bias against SK, which we use as a measure of ideological bias against the capitalistic society of SK. We use a simplified version of the IAT, the Brief Implicit Association Test (\citealp{sriramandgreenwald:2009}; \citealp{noseketal:2014}). This test is to measure the strength of an individual’s unconscious association between categories of NK or SK and words of positive and negative attributes. We select four stimulus images that are clearly identifiable for the categories of NK and SK, as shown in Figure~\ref{fig:stimulus_for_iats}: the national flags of two Koreas, the maps of two Koreas, two Koreas’ uniformed soldiers, and the official full names of the two Koreas--—the Democratic People’s Republic of Korea (NK) and the Republic of Korea (SK). We also select four stimulus words that transparently belong to each of the positive and negative attributes.  


The IAT consists of one practice block and four subsequent blocks. The practice block comprises 20 questions to acquaint subjects with the stimulus items and sorting rules. The four subsequent blocks focus on two contrasting conditions of categories and attributes---``North Korea and good” and ``South Korea and good.” In each block, subjects are required to answer 20 sorting questions. To control the order effect, we randomize the sequence of blocks at the individual level. The information on a pair of categories and attributes is presented at the top of the computer screen, defining the subject’s sorting task in the block. One of the four stimulus images or words appears randomly in the middle of the screen across 20 questions within the block. Upon its appearance on the screen, the subject must quickly figure out whether the stimulus item belongs to either a category or an attribute associated with that block (i.e., presented at the top of the screen). If the stimulus was consistent (resp. inconsistent) with the pair of categories on the top, the subject would hit a key on the right (resp. left) side of the keyboard. This process occurs over a randomized sequence of 20 different stimuli. Sample screens of the IAT and additional details are provided in Online Appendix \textcolor{blue}{A}. 


We follow the psychology literature by constructing the IAT score as the numerical measure of the strength of the implicit association with SK relative to NK. The IAT score is computed for each individual as the difference between the average response time for blocks with the condition of ``South Korea and good” and the average response time for blocks with the condition of ``North Korea and good” divided by the standard deviation of response times across them. As a result, the score ranges between -2 and 2 (\citealp{noseketal:2014}). Note that if a subject is able to sort the stimulus items more quickly when NK is matched with a word of good attribute, the IAT score is positive. It is negative otherwise. Hence, the higher the score is the stronger bias against SK the subject has. 

We use the IAT score as a proxy for ideological bias against the capitalistic society of SK. Given the historical (and still ongoing) ideological conflict between NK and SK as a consequence of the Cold War, it is reasonable to interpret that differences in the IAT score are driven by the extent to which NK refugees are biased against SK for reasons related to their attitudes toward political and economic systems. However, we are also open to other possibilities that the IAT score captures something more personal, such as NK refugees' homesickness and their assimilation experiences in SK. In response to such concerns, we collect detailed individual characteristics, as will be explained in the next subsection and control them in establishing the link between the IAT and market earnings.  


\subsection{Control Variables}
\label{sec-control}

After the subjects completed the market experiments, we conduct a series of cognitive and psychology tests and demographics surveys, through which we try to collect as much information as possible to control for potential confounding factors in the regression analysis. In addition to the sociodemographic information of both subjects, we collect the following extra information. 


First, we collect a set of human capital measurements known to matter for economic outcomes in the market. We use standard Raven’s progressive matrices for cognitive abilities, a Big Five personality test for non-cognitive skills, and ECON 101 and financial literacy tests for economic knowledge. We discuss the details in Online Appendix \textcolor{blue}{C.4.}



Second, we measure each subject’s risk preference. The subjects are asked to make decisions from a series of multiple price lists. \citet{choietal:2021b} provide the full details of the risk preferences experiment. We compute individual-level risk premiums as a measure of risk aversion. 

Finally, for NK refugees, we collect a rich set of variables regarding their life experiences in NK and their assimilation in SK. These include economic class in NK, education level in NK, informal market experience, military service, communist party member, regions from bordering China, birthplace, income, quality of life in SK, and whether they have family members left in NK. We present the summary statistics of the full set of variables in Online Appendix \textcolor{blue}{C.1.}

\section{Empirical Results}
\label{sec4}
\subsection{Validation of IAT Measure}

We begin by validating the IAT score as a measure of ideological bias against a capitalist country. First, we compare NK subjects with the benchmark group of SK subjects. We expect that NK subjects are overall more biased against SK than SK natives. Second, we examine the heterogeneity among NK subjects. Given the presumption that the bias is formed by ideological indoctrination in NK, we expect that the longer people stay in NK, the more likely they hold to a bias against SK. Lastly, we directly compare the IAT measure with the survey responses regarding attitudes toward a market economy and democracy.  

\begin{figure}[htp!]
\small
\begin{center}
\caption{Stimulus and Distributions of IAT}\label{fig:IAT}
\captionsetup{width=1 \textwidth}
\begin{subfigure}[t]{.7\textwidth}
  \centering
  \includegraphics[width=1\linewidth]{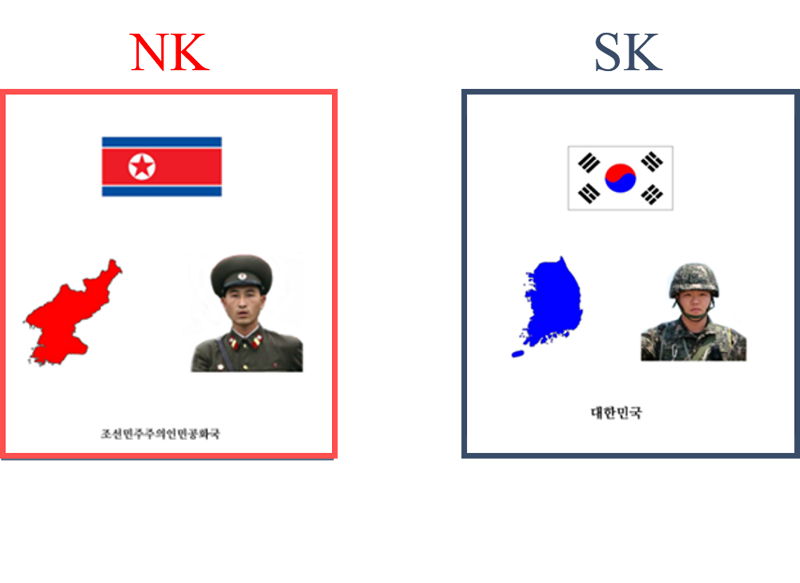}
  \caption{Stimulus}
  \label{fig:stimulus_for_iats}
\end{subfigure}
\begin{minipage}[t]{1\textwidth}
Note (a): The stimulus of the implicit association test includes national flags, maps, images of soldiers, and the official country names--the Democratic People's Republic of Korea (North) and the Republic of Korea (South). \\

\end{minipage}
\begin{subfigure}[t]{.65\textwidth}
  \centering
  \includegraphics[width=1\linewidth]{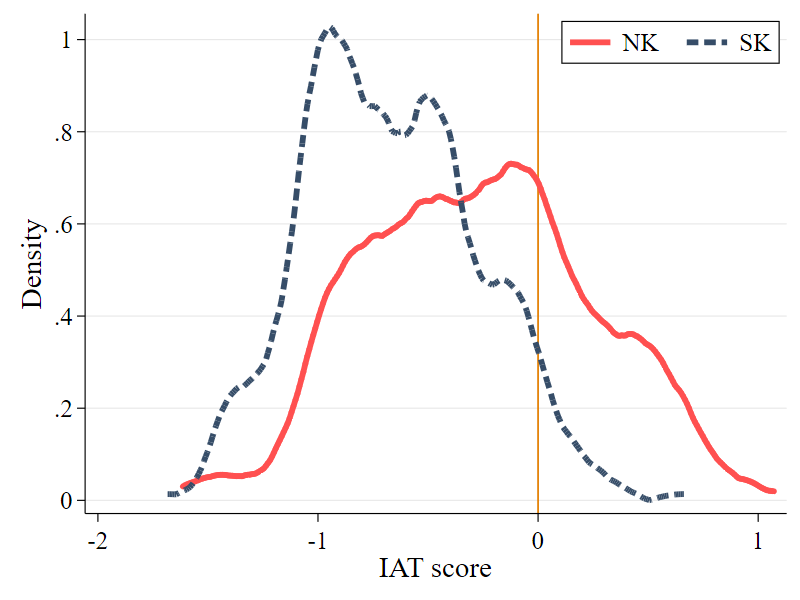}
  \caption{Distribution}
  \label{fig:IAT_distribution}
\end{subfigure}
\end{center}
\begin{minipage}[t]{1\textwidth}
Note (b): The distribution of NK subjects' IAT scores is drawn in solid line. The distribution of NK subjects' IAT scores is drawn in dashed line. They are estimated by Kernel density estimation. The IAT score is 0 (vertical line) if the individual is neutral. A positive IAT score represents a bias against South Korea. \\

\end{minipage}
\end{figure}

\medskip
\textbf{Between-group comparison.} Figure \ref{fig:IAT_distribution} shows the kernel density estimates of the IAT scores for NK and SK subjects. A subject is defined as neutral if the IAT score is 0; biased against SK if the score is positive. We find that while the majority of subjects in both NK and SK groups exhibit pro-SK attitudes, the proportion of pro-NK subjects whose IAT score is positive is 30.3\% among NK subjects, much larger than 4.5\% for SK subjects. On average, NK subjects have higher IAT score than SK subjects: the average IAT scores of NK and SK subjects are -0.27 and -0.69, respectively. The average score gap between NK and SK subjects is 0.42, which is statistically significant ($p$-value$<$0.01). This between-group comparison confirms our first validation check that NK subjects are on average more biased against SK than SK subjects.



\medskip
\textbf{Within-group comparison.} Looking at the within-group variation among NK subjects as in Figure \ref{fig:IAT_distribution}, we examine whether the IAT score is associated with the number of years they spent in NK. We interpret the number of years in NK as a proxy for the degree of exposure to NK's ideological indoctrination against SK. Table~\ref{table:validation_iat_score} presents the regression analysis of the IAT score on the number of years in NK. Column (1) shows the unconditional estimate without any controls, and column (2) with the full set of controls explained in Section \ref{sec-control}.  


The results show that the IAT score is positively correlated to the number of years in NK. According to the point estimate in column (2), one additional year in NK is associated with a 0.013 point increase in the IAT score. To illustrate the magnitude of the estimate, we multiply the estimate (0.013) by the sample average number of years in NK (27.5). The predicted increase in the IAT score amounts to 85\% of the average IAT score gap between NK and SK subjects. 

In addition, we find that the communist party indicator is positively correlated with the IAT score. This is also in line with the presumption that the IAT score captures the degree of ideological indoctrination. We also find NK subjects who belonged to a lower class in NK have a higher IAT score. 

\medskip
Finally, we check the association between the IAT score and explicit attitudes toward market economy and democracy measured by questions borrowed from the World Values Survey. The details of survey questions are provided in Online Appendix \textcolor{blue}{C.3}. We find that, for NK subjects, the correlations of the IAT score with the preferences for a market economy and democracy are significantly negative; -0.14 with the support for a market economy (Pearson correlation test, $p$-value $< 0.01$) and -0.18 with the support for democracy (Pearson correlation test, $p$-value $< 0.01$). This finding further suggests that the variation in the IAT score is driven by the extent to which NK refugees are biased against SK for reasons related to their attitudes to political and economic systems. 

\subsection{IAT and Market Earnings}

We examine the association of ideological bias measured by the IAT score with the profits that subjects earn in market experiments. Our market experiments offer a controlled environment that creates a level playing field for our subjects with different sets of skills and experiences in seeking economic gains. On top of this, we analyze the association between ideological bias and profits by controlling for  potential confounding factors.

Our regression equation is as follows:

\begin{equation}
\label{eq:reg}
    Y_{igr} = \beta \cdot \text{IAT}_{i} + Z_{igr}\gamma + X_{i}\delta + \alpha_{g} + \tau_{r} + \epsilon_{igr} \\   
\end{equation}

\medskip

\noindent where the dependent variable, $Y_{igr}$, is the amount of profits (in the unit of 1,000 KRW) earned by subject $i$ in group $g$ (consisting of eight subjects) in round $r$, $r = 1, ..., 6$. If the subject does not succeed in trading, then his or her profit is zero. $\text{IAT}_i$ is the IAT score of the subject $i$, so $\beta$ is the coefficient of our main interest. $Z_{igr}$ is a vector of buyers’ values and sellers’ costs, which varies randomly within individuals over rounds. $X_{i}$ is a vector of control variables representing individual characteristics ranging from demographic characteristics to human capital measures. We include group- and round-specific fixed effects, $\alpha_{g}$ and $\tau_{r}$ in all specifications. We obtain robust standard errors clustered at the individual level. We estimate the same equation for the DA and BB markets, separately.

Table~\ref{table:DA_profit_iat} presents the results for the DA market (columns (1)-(4) in Panel A) and the BB market (columns (5)-(8) in Panel B), respectively, in the same format. Columns (1) and (5) present the results for all observations with basic sociodemographic controls. Columns (2) and (6) present the results for the markets where both NK and SK subjects participate (mixed markets). To distinguish the association of the IAT score to profits between NK and SK subjects, we interact the IAT score with each of NK and SK indicators. In addition to the basic controls, we further control the set of economic preferences and human capital variables. Columns (3) and (7) present the results for the sample of the NK-only markets with additional controls of variables specific to NK subjects as discussed in Section \ref{sec-control}. Lastly, columns (4) and (8) presents the results for the sample of SK-only markets.

In the DA markets (panel A), we find that NK subjects with a higher IAT score earn less profits. The estimates in columns (1) and (2) are statistically significant at the one percent level. To interpret the economic size of the point estimate, we use the average IAT score gap between NK and SK subjects (0.42) and calculate the predicted effect on profits when the IAT score is increased by the NK-SK IAT gap. The predicted effect is 0.26 in all markets and 0.21 in mixed markets, corresponding to about 4-5\% of the sample median profit (the median profit is 5 for both NK and SK subjects). When we use the sample of the NK-only market, we find that the estimate remains significant and even larger in terms of the magnitude. 

Turning to SK subjects, we find that the IAT score is not significantly associated with profits. In columns (1) and (2), the interaction term of IAT with SK indicator is negative but insignificant. The result for SK subjects in the SK-only market (column (4)) reports basically the same finding. We cannot fully explain why the results are so different between NK and SK subjects. One possible explanation is that the IAT  captures different things for NK and SK subjects. For example, for SK subjects, the IAT possibly captures the attitude of dissidents, which is unlikely to affect profits. 

The results from the BB market (panel B) are similar to those from the DA market; NK subjects with a higher IAT score earn less profits in the BB market as well. Again, the association is not significant among SK subjects. The magnitude of the effect remains similar in NK subjects. For example, if we use the point estimate in column (7) with the sample of the NK-only market, an increase in IAT score by the average gap between NK and SK subjects is associated with a reduction of profits by 7\%. 

\section{Mechanisms}
\label{sec5}

To explain the association between the IAT score and profits for NK subjects, we propose the following hypotheses on expectation and trading behavior. First, NK subjects with a higher IAT score have a lower expectation about profits from the market. Second, the lower expectation could self-fulfill by participating with a lower aspiration level and actually earning lower profits in the market: NK subjects with a higher IAT score engage in trading with a lower ``target profit," that is, are more likely to bid low-profit offers. 

\subsection{Expectation about Profits}

To test the first hypothesis, we examine how the BDM-elicited expectation is related to the IAT score. We run the regression of the IAT score on the expected profit with the full set of control variables we used in our regression for profits and the average amount of profits that each subject earned in the first six rounds before the implementation of BDM.  

Table \ref{table:BDM} presents the results in the same format as in Table~\ref{table:DA_profit_iat}. The results are consistent with our hypothesis. We find that NK subjects with a higher IAT score tend to have lower expectations about profits. Consistent with the results in Table~\ref{table:DA_profit_iat}, the association is only significant for NK subjects. The association for SK subjects is insignificant. It is also worth noting that the effects are larger in magnitude for the expected profits than for the actual profits. For example, using the point estimate in column (3), the effect of an increase in the IAT score by the NK-SK gap is about 17\% of the median profit. 

A natural next question is whether the lower expectation of earnings prior to market participation really affects their behaviors in the market. One potential channel is that subjects with a lower IAT score engage in trading with a lower aspiration of earnings or a lower ``target profits,'' that is, they are more likely to make or accept low-price offers and seek profits less actively. In the next subsection, we investigate the association of IAT score with subjects' trading behavior at the micro level---behaviors of proposing and accepting individual offers. 

\subsection{Target Profits in the Market}

We next turn to trade behavior regarding subjects' target profit in the market experiments to understand the association of the IAT score and profits. We use individuals' bidding data to construct the variable of target profit. Let $b_{ijgr}$ denote the amount of the $j$-th bid that subject $i$ submits in group $g$ and round $r$. Note that the subjects can submit as many bids as they want to. Then, we compute $s_{ijgr} = b_{ijgr} - c_{igr}$ for sellers and $s_{ijgr} = v_{igr} - b_{ijgr}$ for buyers, where $c_{igr}$ and $v_{igr}$ are the cost and value of the commodity for sellers and buyers, respectively. If the bid is accepted by the other participant, the target profit associated with that bid is realized and becomes the subject's profit. We take the average of $s_{ijgr}$ for all $j$'s in round $r$ and use it as a proxy variable for the subject's target profit. 

\begin{figure}[htp!]
\small
\begin{center}
\caption{Target Profit}\label{fig:targetprofit}
\captionsetup{width=1 \textwidth}
\begin{subfigure}{.6\textwidth}
 \centering
  \includegraphics[width=1\linewidth]{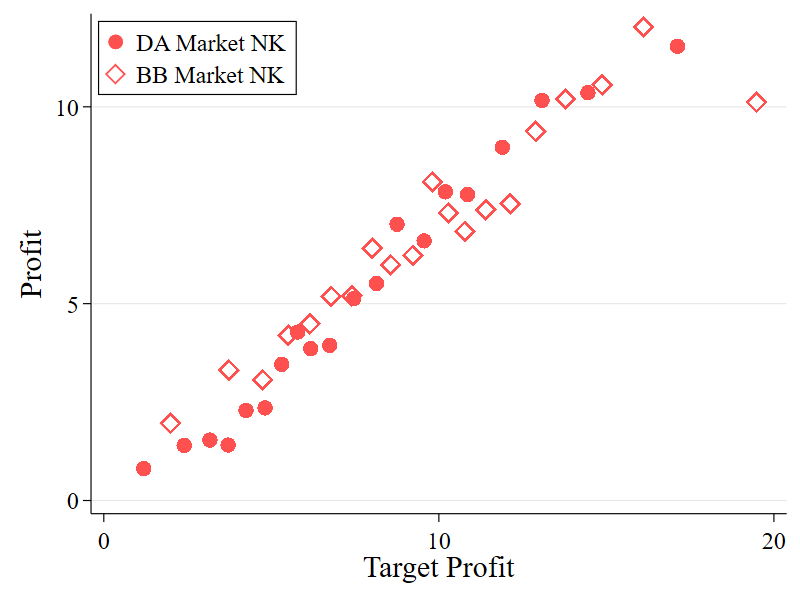}
 \caption{Target Profit and Profit}
 \label{fig:targetprofit_1}

\end{subfigure}
 \begin{minipage}[t]{1\textwidth}
Note (a): We divide the NK sample in each of the DA and BB markets by 20 equal-sized groups depending on the level of their target profits.  On the x-axis (y-axis), we plot the mean of target profit (actual profit) in each group.\\
\end{minipage}
\begin{subfigure}{.6\textwidth}
 \centering
 \includegraphics[width=1\linewidth]{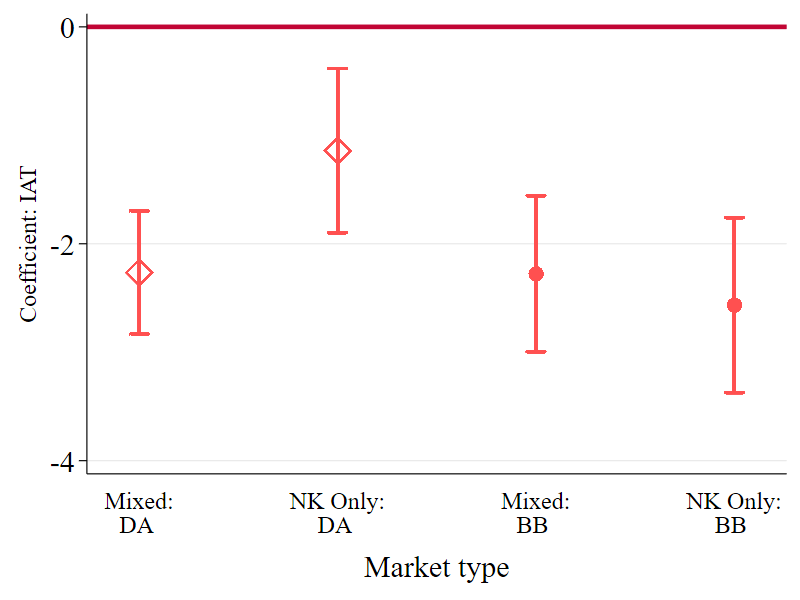}
 \caption{Target Profit and IAT}
  \label{fig:targetprofit_2}
\end{subfigure}
\end{center}
 \begin{minipage}[t]{1\textwidth}
Note (b): We run regression equation~(\ref{eq:reg}) with target profits as the dependent variable and then plot coefficient $\beta$ in each market type. 
In all regressions, we control  gender, age, age squared, marital status, standardized raven score, ECON 101 test, financial literacy, and Big Five personality measures. The full results are presented in Online Appendix B.
\end{minipage}
\end{figure}

Before examining the association between the IAT score and this trading behavior, we examine whether a lower target profit indeed leads to lower profits. Figure \ref{fig:targetprofit_1} shows the relationship between the actual profits and target profits.  In Figure \ref{fig:targetprofit_1}, we divide the NK sample in each of the DA and BB markets by 20 equal-sized groups, depending on the level of their target profits.  Then, we plot the mean of actual profit in each group. We observe a strong positive association between target profits and actual profits for the NK subjects. Online Appendix \textcolor{blue}{C.5} reports the regression results that confirm the patterns in Figure \ref{fig:targetprofit_1}.
It suggests that lower actual profits are driven by subjects' lower aspirations of earnings in trading. This correlation pattern is also found for the SK subjects (see Figure \textcolor{blue}{C.5.1}).

Figure \ref{fig:targetprofit_2} presents the estimated coefficients of IAT and their 95\% confidence intervals for the NK subjects from the regression analysis on subjects' target profits with equation (\ref{eq:reg}). The results are presented for each market type of the DA and BB markets. We observe a strong negative association between IAT and target profits: NK subjects with a higher IAT score engage in trading with a lower target profit. The average target profit is around 7.7 in the DA and 9.6 in the BB markets. 
For instance, in the Mixed DA market (first coefficient), an increase in NK-SK IAT gap decreases target profit by 12.5\%.
All of these coefficients are significant at 5\% level. 
For the SK subjects, we do not find any significant correlation between them (see Figure \textcolor{blue}{C.5.2}).

Online Appendix \textcolor{blue}{C.6} reports that IAT is related to other trading behaviors, such as passivity in trading and the number of bidding which imply other possible mechanisms.

\section{Discussions}
\label{sec6}
Given the historical conflicts between NK and SK as a consequence of the Cold War, we interpret that our measure of IAT score captures NK subjects' implicit bias against the capitalistic society of SK. We show that this bias is associated with negative performance in the market through trading with a low aspiration of earnings. However, we acknowledge other possibilities of explaining our findings and discuss our attempts to rule them out. 

First, the measure of IAT score may capture something other than ideological bias against capitalist SK. For instance, it can capture subjects' unobserved human capital or preferences, which in turn affect economic performance in the market experiment. Alternatively it can reflect NK subjects' difficulties in assimilation into SK for various reasons including homesickness, which are correlated with market earnings. To mitigate these concerns, in the regression analysis, we used an extensive list of NK refugees' characteristics, including a set of human capital variables and their experiences in NK and SK. In this robustness analysis, following the spirit of \citet{altonji2005selection}, we examined whether the estimated association is much affected by the inclusion of controls that, a priori, should be correlated with market earnings. We find that the association of the IAT remains robust to additional correlates of unobserved preferences, human capital, and experiences of assimilating into SK.

Second, the variation of market earnings in the experiment may reflect differences in motivation or preferences regarding effort provision in the artificial environment of the experiment, which may in turn be correlated with the IAT score. To address this concern, we use the number of bidding in a given market as a proxy for motivations of effort provision  and add it in the specifications of Table \ref{table:DA_profit_iat}. We find that the association between the IAT and profits remains robust to the inclusion of this in Online Appendix \textcolor{blue}{C.7}.

Finally, our measure of ideological bias could be too coarse in the sense that it could not pinpoint only the contrast between two economic regimes--capitalism versus socialism. We deliberately chose to use stimulus images that clearly contrast between NK and SK. Because a confrontation between capitalism and socialism is the most salient and important nature of the conflicts between NK and SK, it is plausible to assume that the stimulus images used in the IAT aroused the subjects' attitudes toward economic regimes. Applying stimulus images specific to economic regimes is a potential avenue for future research.



\newpage
\setlength\bibsep{8pt}
\bibliographystyle{ecta}
\bibliography{references_vof}

\newpage

\begin{singlespace}

\newpage

\newpage
\begin{table}[htb]
\centering
\caption{Heterogeneity of IAT score among NK Refugees}
\label{table:validation_iat_score}
\addtolength{\tabcolsep}{14pt}
\begin{tabular}{lcc}
\toprule
                 & (1)                         & (2)                                 \\ \midrule
Years in NK      & 0.009\textsuperscript{***}   & 0.013\textsuperscript{**}   \\
 & (0.003)                       & (0.006)                     \\
Low class in NK &   & 0.183\textsuperscript{***}   \\
&   & (0.066)\\
Communist party &  & 0.306\textsuperscript{**}   \\
& & (0.150)\\
Constant         & -0.524\textsuperscript{***}  & -1.182                     \\
                 & (0.092)  & (0.689) \\ 
\midrule                 
Basic controls           & No                                       & Yes                           \\
Human capital    & No                                             & Yes                           \\
Experience in NK & No                                                 & Yes                           \\
Adaptation in SK & No                                                  & Yes                           \\
Explained NK-SK gap    & 58.9\%                                   & 85.1\%                  \\ 
Observations      & 287         & 287     \\
R-squared      & 0.030     & 0.128  \\ 
\bottomrule
\end{tabular}
\begin{minipage}[t]{1\textwidth}
\small
Note: The dependent variable is the IAT score. Robust standard errors are presented in parentheses. */**/*** represent significance levels 10\%/5\%/1\%. Basic controls include gender, age, age squared, marital status. Human capital control variables include standardized raven score, ECON 101 test, financial literacy, and Big Five personality measures. The variables of experiences in NK include lower economic class in NK, education level in NK, informal market experience, military service, communist party member, regions from bordering with China, and regions from Pyoungyang or Gaesung (big cities in NK). The variables about adaptation in SK include log income in SK, qualify of life in SK and whether they have family members left in NK. The explained NK-SK gap (\%) is the part of the average difference in the IAT score between NK and SK that can be explained by the estimate for Years in NK and the average number of years in NK. The full results are presented in Online Appendix B.
\end{minipage}

\end{table}

\newpage

\begin{table}[htb]
\setlength{\tabcolsep}{14pt}
\small
\centering
\caption{IAT and Profits in Double Auction Markets and Bilateral Bargaining}
\label{table:DA_profit_iat}
\begin{tabular}{lcccc}
\toprule
\multicolumn{5}{c}{A. Double Auction Market} \\
Market compositions & All & Mixed & NK Only & SK Only  \\
 & (1) & (2) &  (3) & (4) \\ \midrule
IAT (against SK) & & &-0.774\textsuperscript{**}&-0.287\\
 & & & (0.391)&(0.289) \\
IAT $\times$ NK & -0.619\textsuperscript{***} & -0.506\textsuperscript{***}   \\
 & (0.211) &(0.246) & &\\
IAT $\times$ SK & -0.185 & -0.157  \\
 & (0.200) &(0.263) & &\\
NK & -0.252 & -0.372  \\
 & (0.256) &(0.296) & &\\
Observations & 3,444 & 1,890 & 768 & 786  \\
R-squared & 0.652 & 0.667 & 0.616 & 0.684\\ \midrule
\multicolumn{5}{c}{B. Bilateral Bargaining} \\
Bargaining compositions & All & Mixed & NK Only & SK Only  \\
 & (5) & (6) &  (7) & (8) \\ \midrule
IAT (against SK) & & &-0.789\textsuperscript{**}&0.600\\
 & & & (0.341)&(0.455) \\
IAT $\times$ NK & -0.719\textsuperscript{**} & -0.851\textsuperscript{**}  \\
 & (0.289) &(0.424) & &\\
IAT $\times$ SK & 0.227& -0.425  \\
 & (0.402) &(0.664) & &\\
NK & -0.069 & 0.037  \\
 & (0.445) & (0.788) & &\\
 Observations & 574 & 315 & 128 & 131  \\
R-squared & 0.256 & 0.283 & 0.535 & 0.377\\ \midrule
Round FE & Yes & Yes  & Yes& Yes   \\
Basic controls & Yes & Yes  & Yes& Yes  \\
Preference & Yes &  Yes &Yes & Yes  \\
Human capital & Yes & Yes &Yes & Yes  \\
Experience in NK &  & & Yes \\
Adaptation in SK &  & & Yes\\ 
\bottomrule
\\[-0.8em]
\end{tabular}
\begin{minipage}{1.0\textwidth}
Note: The dependent variable is the amount of profit (1,000 KRW). Robust standard errors, clustered by individual, are presented in parentheses. */**/*** represent significance levels 10\%/5\%/1\%.
Group- and round-specific fixed effects are included in every specification.
Basic controls include gender, age, age squared, and marital status. Preference controls include seller dummy, buyer's given value and seller's given cost, and risk preferences (average CE). Human capital control variables include standardized raven score, ECON 101 test, financial literacy, and Big Five personality measures. The variables of experiences in NK include lower economic class in NK, education level in NK, informal market experience, military service, communist party member, regions from bordering with China, and regions from Pyoungyang or Gaesung (big cities in NK). The variables about adaptation in SK include log income in SK, qualify of life in SK and whether they have family members left in NK. The full results are presented in Online Appendix B.
\end{minipage}
\end{table}

\newpage
\begin{table}[htb]
\small
\centering
\caption{IAT and Expected Profits in Double Auction Markets}\label{table:BDM}
\begin{tabular}{lcccc}
\toprule
Market compositions & All & Mixed & NK only & SK only \\
 & (1) & (2) &  (3) & (4)  \\ \midrule
IAT (against SK) & & &-2.096\textsuperscript{***}&0.456\\
 & & & (0.598)&(0.496) \\
IAT $\times$ NK & -1.663\textsuperscript{***} & -1.840\textsuperscript{***}  \\
 & (0.300) &(0.366) & &\\
IAT $\times$ SK & 0.465 & 0.261  \\
 & (0.300) &(0.403) & &\\
NK & -0.912\textsuperscript{**} & -0.845\textsuperscript{*}  \\
 & (0.418) &(0.490) & &\\
Observations & 574&315&128&131 \\
R-squared &0.256&0.283&0.535&0.377\\ \midrule
Previous average profit & Yes & Yes  & Yes& Yes \\
Group FE & Yes & Yes  & Yes& Yes  \\
Basic controls & Yes & Yes  & Yes& Yes\\
Preference & Yes &  Yes &Yes & Yes  \\
Human capital & Yes & Yes &Yes & Yes  \\
Experience in NK &  &   & Yes &   \\
Adaptation in SK &  &  & Yes & \\ 
\bottomrule
\\[-0.8em]
\end{tabular}
\begin{minipage}{1.0\textwidth}
Notes: The dependent variable is the amount of expected profits elicited by the BDM method. Robust standard errors, clustered by individual, are presented in parentheses.
 */**/*** represent significance levels 10\%/5\%/1\%.
Group fixed effects and  individual level average profit in the previous 6 rounds are controlled in all specification.
Basic controls include gender, age, age squared, marital status. Preference controls include seller dummy, buyer's given value and seller's given cost, and risk preferences (average CE). Human capital control variables include standardized raven score, ECON 101 test, financial literacy, and Big Five personality measures. The variables of experiences in NK include lower economic class in NK, education level in NK, informal market experience, military service, communist party member, regions from bordering with China, and regions from Pyoungyang or Gaesung (big cities in NK). The variables about adaptation in SK include log income in SK, qualify of life in SK and whether they have family members left in NK. The full results are presented in Online Appendix B.
\end{minipage}
\end{table}

\end{singlespace}

\clearpage
\newpage



\section*{Supplementary material}

# Online Appendix A. Instruction

## Online Appendix A.1. Experiment

Thank you for your participation.

We got IRB approval from Seoul Nation University. IRB No. 1606/002-003, Date 2016.6.21

You are not allowed to talk to others during the experiment, except to ask the experimenter a question. Please turn off your cell phone.

**Outline of the experiment:**

This experiment consists of three parts. You will be briefed before each stage of the experiment. At the end of this experiment, you will receive a cash reward. Details of determining the reward amount differ for each part of the experiment, so please ensure you understand the following explanation of each parts.

**Earnings in the experiment:**

Your total remuneration for today will be determined by the sum of your earnings in each stage, in addition to the basic participation reward of KRW45,000.

## Part 1.

In the first stage of the experiment, you will be assigned the role of either a seller or a buyer. In a market with both buyers and sellers, the sellers and buyers will earn money by selling or buying an imaginary item.

### Formation of market

Eight participants in this room get to form a single market. Half of the participants in each market get to play the role of seller, and other half get to play the role of buyer. The formation of groups and roles are assigned randomly by a computer. The market will open seven times in total. Market members and their roles will remain constant throughout the experiment.

### Role of seller:

Please note that the item is imaginary. Selling the item incurs a certain cost on the seller. The selling cost of the item will be assigned randomly in each market. You can propose the price of the item, presenting the price you want to sell to others in the same group. You can sell your item within three minutes. You cannot sell for less than the randomly given selling cost. Your final earning in the market will equal the price at which you actually sold the item minus the selling cost. If no buyer buys your item, you earn nothing.

Example 1-1

When seller has a cost of KRW 2000. He sold his item at the price of KRW 18000. Then his profit in this market is KRW 16000.

Example 1-2

When seller has a cost of KRW 6000. He did not sell item. Then his profit in this market is KRW 0.

### Role of buyer:

For buyer, the item has a value. The value will be assigned a randomly generated value at the time of market opening. You can buy the item how much you will pay for the item, present your bid to others, and can buy the item within three minutes. You cannot buy at a higher price than the value you have. Your final earning will equal your purchase value minus the price at which you actually bought the item. If you did not manage to buy the item, you earn nothing.

Example 2-1

When buyer has a value of KRW 25000. He bought his item at the price of KRW 14000. Then his profit in this market is KRW 11000.

Example 2-2

When buyer has a value of KRW 17000. He did not buy the item. Then his profit in this market is KRW 0.

**Transaction:**

The market will open seven (7) times excluding the two (2) practice openings. Each market opening lasts three (3) minutes (180 seconds). You can make a transaction within three (3) minutes in order to earn money. Once you make a transaction, you cannot make any more transactions in the remaining time.

**Earnings in Part 1:**

The amount of money you earned in each market opening will be presented at the end of each market. Your total earnings from market transactions will be determined by randomly drawing the results of two out of seven market.

**Structure of the screen:**

Figure 1 shows the screen shot for the screen you play. Please follow the number and check functions of buttons. As we noted, there are 2 practice markets for your understanding of the game.

1. *Your role as either a seller or a buyer will be displayed publicly.*

2. *The bids of all the buyers in your market will be displayed publicly.*

3. *The prices set by all the sellers in your market will be displayed publicly.*

4. *The list of completed transactions in your market will be displayed publicly. Four sets of sellers and buyers get to transact one-to-one.*

5. *The remaining time is displayed at the top of the screen. You cannot make a tran saction if you fail to do so within the allotted time.*

6. *Among the bids from different buyers, click on the one that you wish to trade for. The bids will be listed in descending order of price.*

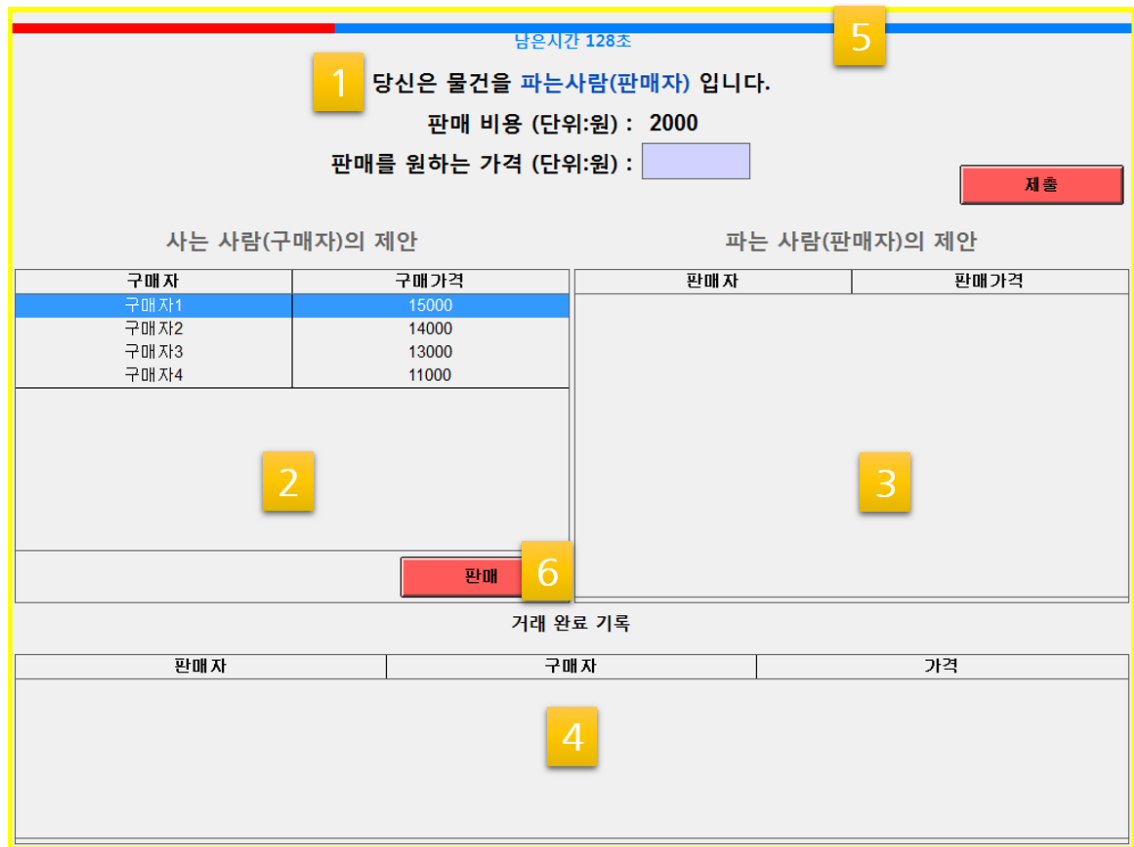

Figure A.1. Decision screen

We will now begin the two practice markets and following seven actual markets. As we noted, your final earnings from market transactions will be determined by randomly drawing the results of two out of seven market openings.

**We remind you again the <u>5 important information</u> in Part 1.**

1. First of all, confirm your role (seller or buyer) and corresponding given value or cost. Your role during practice will remain unchanged throughout the actual experiment.

2. There are four 4 sellers and four 4 buyers in your market.

3. Each market remains open for three minutes (180 seconds). You cannot take part in any transactions after the market closes.

4. You may not be able to make a transaction with some probability. You won't be able to make a transaction if you are left with a seller with a higher selling cost than your purchase value, or a buyer with a lower purchase value than your selling cost.

5. Both the seller and the buyer will be excluded from the market once they have made a transaction. Once you complete a transaction, you will no longer be able to

participate in the market. Those who already completed their transactions will be out from the market while the rest continues to transact in the market.

## Part II.

In the second stage of the experiment, your role will be remained same in Part 1. However, in a market there are one seller and one buyer.

### Formation of market

You will be matched randomly whom you are matched in a same group in Part 1. In a market with both buyer and seller, the seller and buyer will earn money by selling or buying an imaginary item to the other. The market will open three times in total.

### Role of seller:

Please note that the item is imaginary. Selling the item incurs a certain cost on the seller. The selling cost of the item will be assigned randomly in each market. You can propose the price of the item, presenting the price you want to sell. You can sell your item within 90 seconds. You cannot sell for less than the randomly given selling cost. Your final earning in the market will equal the price at which you actually sold the item minus the selling cost. If no buyer buys your item, you earn nothing.

### Role of buyer:

For buyer, the item has a value. The value will be assigned a randomly generated value at the time of market opening. You can buy the item how much you will pay for the item, present your bid to the seller, and can buy the item within 90 seconds. You cannot buy at a higher price than the value you have. Your final earning will equal your purchase value minus the price at which you actually bought the item. If you did not manage to buy the item, you earn nothing.

### Transaction:

Different from Part 1, the transaction can happen in any case which means the cost of the seller is always lower than the value of the buyer. Each market opening lasts 90 seconds. Once you make a transaction, you cannot make any more transactions in the remaining time.

### Earnings in Part 2:

The amount of money you earned in each market opening will be presented at the end of each market. Your total earnings from market transactions will be determined by randomly drawing the results of one out of three market.

**We remind you again the <u>4 important information</u> in Part 2.**

1. Your role (seller or buyer) in Part 2 will remain unchanged and same with Part 1.

2. You will be randomly matched the one who belongs to the same group in Part 1.

3. The cost of seller is always lower than the value of buyer which means that trans action could be happen.
4. The market will be opened for 90 seconds.

**Part III.**

During this experiment you will see two lotteries, named Lottery L (left) and Lottery R (right), on the top of each page. You can choose the lottery what you want between L and R. If you choose Lottery L, then you can get KRW 8000 with some probability. Otherwise, KRW 0. If you choose Lottery R, then you can get a certain amount of money which is less than KRW 8000. There are 40 pairs of questions in total. Therefore, you will make a decision for a question after you compare the two lotteries.

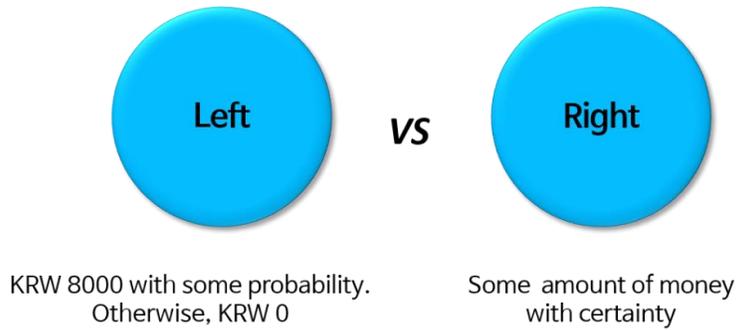

KRW 8000 with some probability.     Some amount of money
Otherwise, KRW 0                     with certainty

**Probability:**

There are five variations in probability: 5%, 25%, 50%, 75%, 95%. You should imagine that, there are 100 balls in the black box. The balls should be red or blue. When you draw a ball from the box, you can earn 8000 KRW when the ball is blue ball. If the ball is red ball, you can earn nothing.

The concept of 5% for the winning KRW 8000 means that you draw a ball from black box where 5 blue balls and 95 red balls exist in the black box.

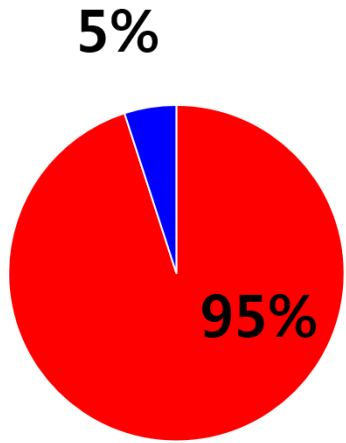

The concept of 25% for the winning KRW 8000 means that you draw a ball from black box where 25 blue balls and 75 red balls exist in the black box.

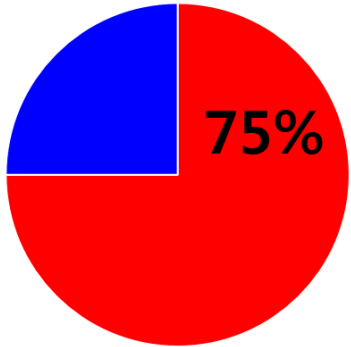

The concept of 50% for the winning KRW 8000 means that you draw a ball from black box where 50 blue balls and 50 red balls exist in the black box.

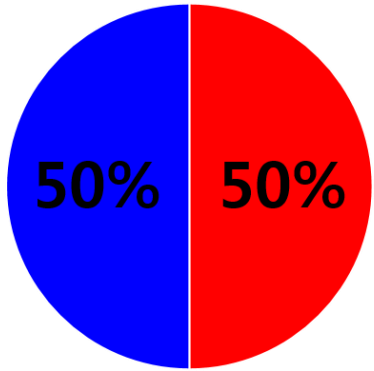

The concept of 75% for the winning KRW 8000 means that you draw a ball from black box where 75 blue balls and 25 red balls exist in the black box.

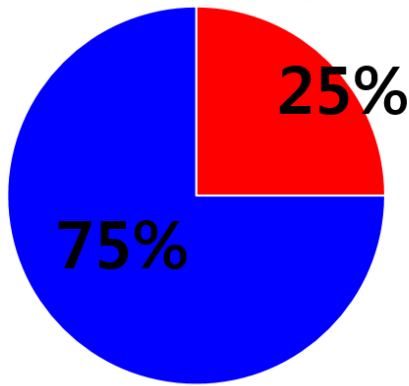

The concept of 95% for the winning KRW 8000 means that you draw a ball from black box where 95 blue balls and 5 red balls exist in the black box.

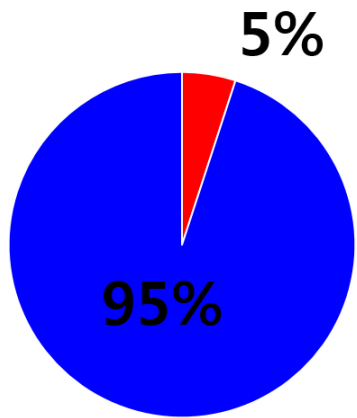

**Earnings in Part 3:**

One out forty question will be randomly selected after you finish the experiment. Depending on your choice of that question. If you chose Lottery L in the selected question, the computer will draw a lottery and will let you know the amount of money you will get in this part. If you chose Lottery R in the selected question, you can get the certain amount.

# Online Appendix A.2. Implicit Association Test

From now on, you participate in a sorting task. There are figures and words which belong to NK or SK. These are possible figures you will see in the sorting task.
.

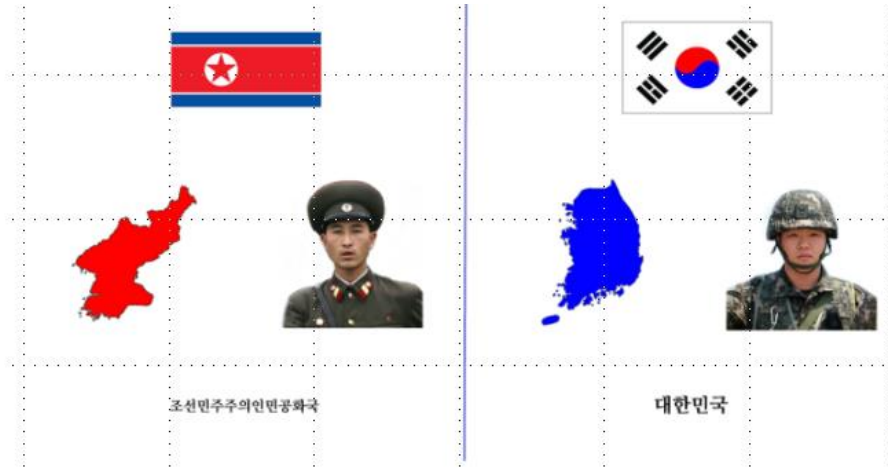

These are the set of words you will see in the sorting task.

| Good | Bad |
| --- | --- |
| 호탕하다<br>Generous or Forgiving | 뒤끝있다<br>Grudging |
| 품위있다<br>Elegant | 겉과 속이 다르다<br>Deceitful |
| 똑똑하다<br>Smart | 어리석다<br>Stupid |
| 신뢰가 간다<br>Trustworthy | 간사하다<br>Sly |

When you participate in the sorting game, there is a condition on the top of the screen. The condition is a combination of North Korea/South Korea or Good/Bad.

There will be a figure or a word in the center of the screen, so please sort this figure or word whether it is consistent with the condition or not.

When doing a sorting task you are asked to quickly report the word or figure is consistent or inconsistent with the condition. If you think the word or figure is consistent, then click "i" key if it is inconsistent please click "e" key.

For example, in the below case, the condition is "Good or North Korea". The figure is North Korea, so you should click "i" key because it is consistent with the condition.

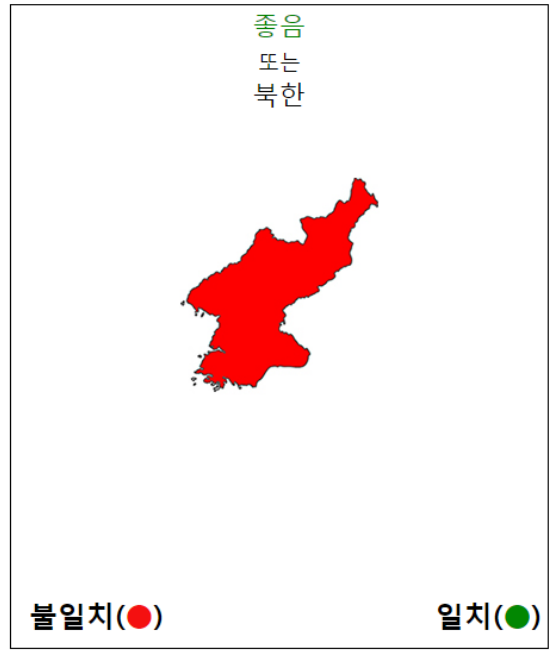

You will participate in 20 practice sorting tasks and then you will participate in 80 sorting tasks. The condition on the top of screen will be changed after every 20 rounds (block). We will notify you when the condition is changed.

# Online Appendix A.3. Construction of Implicit Association Test Score.

We use the following restrictions when we construct IAT score. Each individual has 80 response time and whether the answer is correct or not.

1. Remove sorting task faster than 10 milliseconds

2. Remove first four tasks of each response block

3. Recode <400 ms to 400 ms and >2000 ms to 2000 ms

4. Compute IAT score separately for each pair of two consecutive blocks separately and then average

5. Remove block with more than 10% fast responses

6. Remove block with more than 30% of wrong answers.

# Online Appendix B. Full Table Results

## Online Appendix B.1. Full Table 1

Table B.1. Full table for Table 1

| VARIABLES | (1) IAT score (against SK) | (2) IAT score (against SK) |
|---|---|---|
| Years in NK | 0.009*** | 0.013** |
|  | (0.003) | (0.007) |
| Female |  | -0.045 |
|  |  | (0.092) |
| Age |  | 0.031 |
|  |  | (0.035) |
| Age squared |  | -0.001 |
|  |  | (0.000) |
| Married |  | -0.017 |
|  |  | (0.077) |
| Standardized Raven score |  | -0.069* |
|  |  | (0.042) |
| Basic economics knowledge |  | -0.009 |
|  |  | (0.021) |
| Financial Literacy |  | 0.004 |
|  |  | (0.033) |
| Big5: Extraversion |  | -0.011 |
|  |  | (0.009) |
| Big5: Neuroticism |  | -0.007 |
|  |  | (0.007) |
| Big5: Agreeableness |  | 0.005 |
|  |  | (0.010) |
| Big5: Consciousness |  | -0.013 |
|  |  | (0.012) |
| Big5: Openness |  | 0.000 |
|  |  | (0.012) |
| Low class in NK |  | 0.183*** |
|  |  | (0.066) |
| Secondary Education |  | -0.102 |
|  |  | (0.146) |
| Military service in NK |  | -0.232 |
|  |  | (0.149) |
| Member of labor party |  | 0.306** |
|  |  | (0.150) |
| From NK Border Province |  | 0.127 |
|  |  | (0.091) |
| From NK Captial Region |  | 0.200 |
|  |  | (0.142) |
| Left family in NK |  | -0.069 |

|  |  |  |
|---|---:|---:|
|  |  | (0.072) |
| Log income in SK |  | 0.042 |
|  |  | (0.049) |
| Quality of Life - After escape |  | 0.045 |
|  |  | (0.030) |
| Constant | -0.524*** | -1.025 |
|  | (0.092) | (0.694) |
| Observations | 287 | 287 |
| R-squared | 0.030 | 0.128 |

# Online Appendix B.2. Full Table 2 Panel A.

Table B.2. Full table for Table 2 Panel A

| VARIABLES | (1) Profits | (2) Profits | (3) Profits | (4) Profits |
|---|---|---|---|---|
| IAT score (against SK) | | | -0.774** | -0.287 |
| | | | (0.391) | (0.289) |
| NK x IAT | -0.619*** | -0.506** | | |
| | (0.211) | (0.246) | | |
| SK x IAT | -0.185 | -0.157 | | |
| | (0.200) | (0.263) | | |
| NK | -0.252 | -0.372 | | |
| | (0.256) | (0.296) | | |
| Female | 0.084 | -0.028 | 0.755 | 0.067 |
| | (0.159) | (0.211) | (0.476) | (0.290) |
| Age | 0.122** | 0.204*** | 0.163 | -0.119 |
| | (0.055) | (0.068) | (0.158) | (0.125) |
| Age squared | -0.002*** | -0.003*** | -0.002 | 0.001 |
| | (0.001) | (0.001) | (0.002) | (0.002) |
| Married | -0.137 | 0.087 | -0.460 | -0.665** |
| | (0.150) | (0.204) | (0.349) | (0.324) |
| Seller | -0.295** | -0.221 | -0.701** | -0.195 |
| | (0.116) | (0.156) | (0.270) | (0.238) |
| Standardized Raven score | 0.105 | 0.061 | 0.284 | 0.157 |
| | (0.099) | (0.121) | (0.219) | (0.280) |
| Basic economics knowledge | 0.085* | 0.055 | 0.062 | 0.088 |
| | (0.045) | (0.053) | (0.120) | (0.112) |
| Financial Literacy | 0.122 | 0.074 | 0.360** | -0.175 |
| | (0.080) | (0.104) | (0.155) | (0.196) |
| Big5: Extraversion | 0.008 | 0.022 | 0.050 | -0.046 |
| | (0.016) | (0.020) | (0.050) | (0.032) |
| Big5: Neuroticism | 0.018 | 0.041** | -0.003 | -0.007 |
| | (0.014) | (0.019) | (0.034) | (0.024) |
| Big5: Agreeableness | -0.011 | 0.028 | -0.063 | -0.021 |
| | (0.021) | (0.027) | (0.053) | (0.039) |
| Big5: Consciousness | 0.006 | 0.013 | 0.010 | -0.009 |
| | (0.021) | (0.028) | (0.047) | (0.045) |
| Big5: Openness | -0.018 | -0.040 | -0.015 | -0.010 |
| | (0.021) | (0.030) | (0.056) | (0.036) |
| Secondary Education in NK | | | -0.082 | |
| | | | (0.604) | |
| Post secondary Edcation in NK | | | -0.088 | |
| | | | (0.814) | |
| Informal market experience | | | -0.025 | |
| | | | (0.355) | |
| Military service in NK | | | 0.833 | |
| | | | (0.747) | |

| | | | | |
|---|---|---|---|---|
| From NK Captial Region | | | -0.718 | |
| | | | (0.691) | |
| Member of labor party | | | -0.960 | |
| | | | (0.963) | |
| Border Province | | | 0.187 | |
| | | | (0.423) | |
| Left family in NK | | | -0.039 | |
| | | | (0.365) | |
| Log income in SK | | | 0.506* | |
| | | | (0.262) | |
| Quality of Life - After escape | | | 0.121 | |
| | | | (0.139) | |
| Constant | 7.404*** | 6.536*** | 4.617 | 13.691*** |
| | (1.235) | (1.601) | (3.126) | (2.560) |
| | | | | |
| Observations | 3,444 | 1,890 | 768 | 786 |
| R-squared | 0.652 | 0.667 | 0.616 | 0.684 |

# Online Appendix B.3. Full Table 2 Panel B.

| VARIABLES | (1) Profits | (2) Profits | (3) Profits | (4) Profits |
|---|---|---|---|---|
| IAT score (against SK) | | | -0.789** | 0.600 |
| | | | (0.341) | (0.455) |
| NK x IAT | -0.719** | -0.851** | | |
| | (0.289) | (0.424) | | |
| SK x IAT | 0.227 | -0.425 | | |
| | (0.402) | (0.664) | | |
| NK | -0.069 | 0.037 | | |
| | (0.445) | (0.788) | | |
| Female | -0.100 | -0.663 | -0.339 | 0.809** |
| | (0.245) | (0.439) | (0.467) | (0.406) |
| Age | 0.020 | -0.053 | 0.144 | 0.064 |
| | (0.089) | (0.146) | (0.182) | (0.133) |
| Age squared | -0.000 | 0.000 | -0.001 | -0.001 |
| | (0.001) | (0.002) | (0.002) | (0.002) |
| Married | -0.291 | 0.418 | -0.489 | -0.700 |
| | (0.246) | (0.433) | (0.406) | (0.529) |
| Seller dummy | -0.710*** | -0.605 | -0.582 | -0.839** |
| | (0.219) | (0.376) | (0.362) | (0.365) |
| WTP for lotteries | 0.029 | 0.032 | 0.003 | 0.144 |
| | (0.084) | (0.152) | (0.118) | (0.159) |
| Standardized Raven score | 0.273 | -0.040 | 0.238 | 0.509 |
| | (0.186) | (0.343) | (0.237) | (0.413) |
| Basic economics knowledge | 0.123 | 0.250** | 0.205* | 0.077 |
| | (0.079) | (0.119) | (0.122) | (0.153) |
| Financial Literacy | 0.178 | 0.151 | 0.184 | 0.309 |
| | (0.138) | (0.247) | (0.207) | (0.240) |
| Big5: Extraversion | 0.019 | -0.027 | 0.030 | 0.045 |
| | (0.029) | (0.055) | (0.048) | (0.046) |
| Big5: Neuroticism | 0.017 | 0.031 | 0.015 | -0.012 |
| | (0.021) | (0.041) | (0.035) | (0.035) |
| Big5: Agreeableness | 0.010 | 0.055 | 0.034 | -0.025 |
| | (0.034) | (0.059) | (0.057) | (0.056) |
| Big5: Consciousness | -0.018 | 0.011 | -0.026 | -0.054 |
| | (0.035) | (0.065) | (0.062) | (0.050) |
| Big5: Openness | -0.005 | 0.011 | -0.023 | -0.035 |
| | (0.039) | (0.077) | (0.063) | (0.060) |
| Secondary Education in NK | | | -1.228 | |
| | | | (0.855) | |
| Post secondary Education in NK | | | -1.271 | |
| | | | (0.800) | |
| Informal market experience | | | -0.684** | |
| | | | (0.342) | |

| | | | | |
|---|---|---|---|---|
| Military service in NK | | | 0.727 | |
| | | | (0.850) | |
| From NK Capital Region | | | 1.695** | |
| | | | (0.750) | |
| Member of labor party | | | -0.735 | |
| | | | (0.880) | |
| Border Province | | | 1.090*** | |
| | | | (0.389) | |
| Left family in NK | | | -0.016 | |
| | | | (0.373) | |
| Log income in SK | | | 0.372 | |
| | | | (0.245) | |
| Quality of Life - After escape | | | -0.034 | |
| | | | (0.183) | |
| Constant | 8.833*** | 9.341*** | 4.168 | 8.742*** |
| | (1.898) | (2.886) | (3.855) | (3.159) |
| | | | | |
| Observations | 1,722 | 442 | 640 | 640 |
| R-squared | 0.304 | 0.344 | 0.357 | 0.274 |

# Online Appendix B.4. Full Table 3.

| VARIABLES | (1) bdm (1 000 KRW) | (2) bdm (1 000 KRW) | (3) bdm (1 000 KRW) | (4) bdm (1 000 KRW) |
|---|---|---|---|---|
| IAT score (against SK) | | | -2.096*** | 0.456 |
| | | | (0.598) | (0.496) |
| NK x IAT | -1.663*** | -1.840*** | | |
| | (0.300) | (0.366) | | |
| SK x IAT | 0.465 | 0.261 | | |
| | (0.300) | (0.403) | | |
| NK | -0.912** | -0.845* | | |
| | (0.418) | (0.490) | | |
| Female | -0.043 | -0.015 | 0.992 | -0.013 |
| | (0.277) | (0.354) | (0.902) | (0.548) |
| Age | 0.123 | 0.102 | 0.382 | 0.168 |
| | (0.087) | (0.114) | (0.334) | (0.188) |
| Age squared | -0.001 | -0.001 | -0.004 | -0.002 |
| | (0.001) | (0.001) | (0.004) | (0.002) |
| Married | 0.299 | -0.031 | 0.922 | 0.292 |
| | (0.254) | (0.334) | (0.641) | (0.577) |
| Seller | 0.339* | 0.580** | -0.244 | -0.012 |
| | (0.196) | (0.264) | (0.459) | (0.334) |
| WTP for lotteries | 0.353*** | 0.239* | 0.506*** | 0.648*** |
| | (0.082) | (0.124) | (0.160) | (0.151) |
| Profit Last 6 Rounds | 0.161*** | 0.184*** | 0.177* | 0.122 |
| | (0.045) | (0.060) | (0.106) | (0.092) |
| Standardized Raven score | 0.105 | -0.033 | -0.005 | 0.389 |
| | (0.163) | (0.201) | (0.369) | (0.432) |
| Basic economics knowledge | 0.086 | 0.107 | 0.487** | -0.287* |
| | (0.080) | (0.095) | (0.190) | (0.156) |
| Financial Literacy | 0.110 | 0.308* | -0.139 | 0.374 |
| | (0.131) | (0.177) | (0.312) | (0.247) |
| Big5: Extraversion | -0.016 | -0.024 | 0.184** | -0.132*** |
| | (0.033) | (0.043) | (0.090) | (0.048) |
| Big5: Neuroticism | -0.026 | -0.058* | -0.058 | 0.008 |
| | (0.020) | (0.031) | (0.046) | (0.041) |
| Big5: Agreeableness | -0.031 | -0.050 | -0.046 | -0.029 |
| | (0.033) | (0.046) | (0.086) | (0.057) |
| Big5: Consciousness | -0.024 | -0.011 | -0.052 | -0.005 |
| | (0.033) | (0.043) | (0.079) | (0.059) |
| Big5: Openness | 0.047 | 0.055 | -0.030 | 0.121* |
| | (0.039) | (0.061) | (0.098) | (0.062) |
| Secondary Education in NK | | | 0.720 | |
| | | | (1.928) | |
| Post Secondary Education in NK | | | -1.369 | |
| | | | (2.143) | |

| | | | | |
|---|---|---|---|---|
| Informal market experience | | | -0.607 | |
| | | | (0.573) | |
| Military service in NK | | | 3.273*** | |
| | | | (1.108) | |
| From Capital Region | | | 3.666*** | |
| | | | (1.303) | |
| Member of labor party | | | -0.363 | |
| | | | (1.270) | |
| From Border Province | | | 1.622** | |
| | | | (0.745) | |
| Left family in NK | | | -1.410** | |
| | | | (0.621) | |
| Log income in SK | | | 0.179 | |
| | | | (0.380) | |
| Quality of Life - After escape | | | 0.125 | |
| | | | (0.272) | |
| Constant | 0.988 | 0.603 | -9.608 | 0.115 |
| | (1.999) | (2.860) | (6.090) | (3.643) |
| | | | | |
| Observations | 574 | 315 | 128 | 131 |
| R-squared | 0.274 | 0.305 | 0.551 | 0.390 |

# Online Appendix C. Supplementary Results

## Online Appendix C.1. Basic Demographics

In this section, we first compare basic demographics of NK and SK. Below Table shows the demographics of NK and SK samples. Table C.1 shows the basic demographic characteristics of our subjects. As explained, we recruited SK natives to match the age and gender composition of NK refugees. Therefore, between NK and SK subjects, age and gender are well balanced. The average age is 39 and the proportion of female participants is about 70% in both groups. We note that according to official statistics for the population of NK refugees in SK, NK refugees are predominantly females. However, we find that household income is much higher among South Koreans. It is not surprising that North Koreans economically underperform, in part because a major part of their human capital obtained in NK is obsolete in SK. In addition, SK subjects are more educated than NK subjects, although it is hard to compare two countries directly in terms of education because their education systems are fundamentally different. However, note that we do not attempt to compare NK and SK subjects in the current study but exploit the variation within NK subjects in terms of ideological bias. Therefore, the differences between NK and SK subjects are not serious concerns in the current study. We also report various experience in NK for NK subjects.

Table C.1. Basic Demographics

| Sample | NK | | SK | |
|---|---|---|---|---|
| | Mean | SD | Mean | SD |
| Age | 39.077 | 9.263 | 38.718 | 10.101 |
| Female | 0.711 | 0.454 | 0.686 | 0.465 |
| Married | 0.338 | 0.474 | 0.544 | 0.499 |
| Monthly household income | 180.8 | 240.2 | 548.5 | 249.1 |
| Secondary education | 0.526 | 0.5 | 0.227 | 0.42 |
| Postsecondary education | 0.282 | 0.451 | 0.282 | 0.42 |
| **NK experience variables** | | | | |
| Years in NK | 27.48 | 9.552 | | |
| Years in SK | 7.289 | 3.482 | | |
| Years in a third country | 4.306 | 3.667 | | |
| Border provinces | 0.805 | 0.397 | | |
| Pyoungyang or Gaesung | 0.035 | 0.184 | | |
| Lower economic class in NK | 0.436 | 0.497 | | |
| Informal market activities | 0.331 | 0.471 | | |
| Family members left in NK | 0.711 | 0.454 | | |
| Communist party member | 0.077 | 0.267 | | |
| Military service in NK | 0.115 | 0.32 | | |
| # of subjects | 287 | | 287 | |

Note: The unit of monthly household income is 10,000 KRW (about 10 USD). The parenthesis shows the standard deviation

of the variable.

## Online Appendix C.2. Balance test

As we mentioned, NK is divided into two groups: one with NK only market and the other one with mix market. Table C.2 tests the balance of our NK samples in each market. We note that our NK sample in NK only market are on average 2 years younger compared to Mix market NK and this difference is significant at 5% level. Except for the age variable, other demographics are not significantly different at the 5% level which shows our sample is well-balanced between markets. Especially, we note that our IAT score is not significantly different between the market structure. Fig C.2. shows the distribution of age in each market.

Table C.2. Balance check of NKs between markets

|  | NK only | | Mixed market | | | |
| --- | --- | --- | --- | --- | --- | --- |
|  | Mean | SD | Mean | SD | T-statistics | p-value |
| IAT score | -0.30 | 0.50 | -0.23 | 0.53 | 1.05 | 0.295 |
| Age | 38.07 | 9.38 | 40.31 | 9.00 | 2.05 | 0.041 |
| Female | 0.71 | 0.46 | 0.71 | 0.45 | 0.08 | 0.936 |
| Married | 0.38 | 0.49 | 0.29 | 0.45 | -1.66 | 0.098 |
| Monthly household income | 192.12 | 291.07 | 166.88 | 156.89 | -0.89 | 0.377 |
| Secondary education | 0.52 | 0.50 | 0.53 | 0.50 | 0.27 | 0.789 |
| Post-secondary education | 0.25 | 0.44 | 0.32 | 0.47 | 1.21 | 0.227 |
| # of subjects | 128 | | 159 | | | |

Note: We note that age variable is significantly different at 5% level between the NK only and Mix market. The unit of Monthly household income is 10,000 KRW.

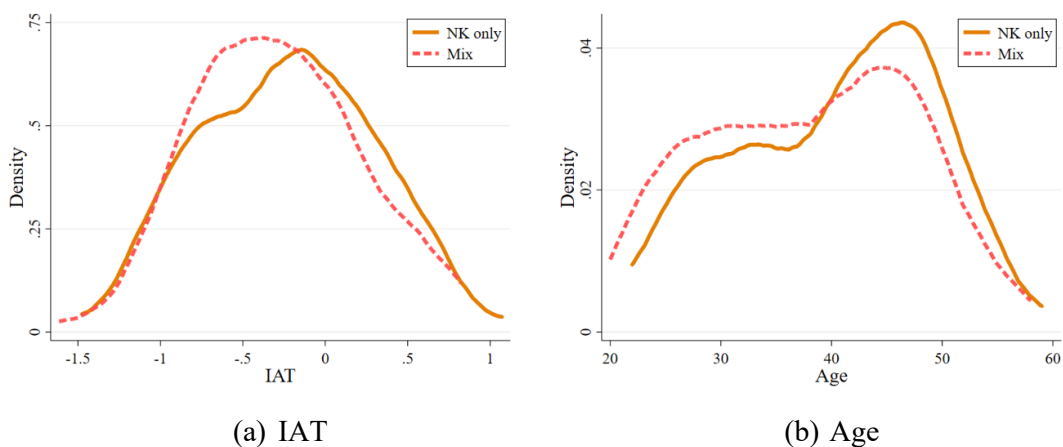

(a) IAT           (b) Age

Fig C.2. Distribution of IAT and Age

## Online Appendix C.3. Explicit Attitudes

In this section, we explain how we construct our explicit attitudes measures. For the preference of market economy we had following Liker scale survey questions (i) performance based incentives, (ii) market rules, (iii) private ownership, (iv) market competitions. As for performance-based incentives, the following question is presented: "One should be paid a higher wage than another, if the former works better than the latter, even though they are of the same age and same rank in the same company." Then, as for market rules, we use the sentence "Capitalism is a system that deems any means justifiable as long as it makes more money." Regarding private ownership, we present subjects "It is better for the state rather than individuals to own firms, lands, residences, etc." Finally, about the market competition, we presented two questions: "Competition among individuals is necessary for economic development" and "It is more convenient to live in a collectivist society without competition." The range of each Liker scale question is from 1 (Strongly disagree) to 5 (Strongly agree). The preference for market is the sum of these questions.

      We also ask questions about five different aspects of democracy—--(i) the multi-party political system, (ii) freedom to vote, (iii) individualism (vs. collectivism), (iv) anti-strong political leader, and (v) human equality. The question for the multiparty system is "Multiple political parties are necessary to aggregate diverse opinions." Regarding freedom of voting, the following two questions are asked: "national leaders should be selected among multiple candidates by people's free will" and "I make my own voting decision rather than following others opinions." As for individualism, "One's liberty can be sacrificed for the benefit of the whole" asked. The question for anti-strong political leaders is "A strong leader is necessary for national development.". Lastly, the question for human equality is " man should be treated equally regardless of his income, place of origin, educational attainment, ethnicity, or nationality." Again, we use the Likert scale ranges from 1 to 5 in each questions and summing up to construct the measure.

# Online Appendix C.4. Human Capital Measurement

In this section, we summarize our human capital measurement. First, we measure cognitive and non-cognitive skills. Second, we also measure individual differences in economic knowledge and skills, we conduct simple ECON 101 and financial literacy tests to measure each subject's level of knowledge about a market economy. In the ECON 101 test, we basically test whether they understand how changes in supply and demand affect the market price using seven questions. For example, one question is as follows: ``Would the market price increase as demand increases in the market?" These questions are designed to test whether they have basic knowledge about the market's price-adjustment mechanism. The financial literacy test is designed to test whether they understand the basic concepts of the interest rate, economic growth, and returns from savings.

Table C.4. shows the results for human capital measurement. We find that there is a large difference in cognitive ability between NK and SK subjects. It is beyond the scope of the current study to account for the difference, but the cognitive gap might exist because North Korean subjects experienced nutritional insufficiency in their childhood or their early life conditions were not favorable. Unlike the case of cognitive ability, all the measures of non-cognitive skills are not significantly different at the 5% level.

Table C.4. Human Capital Measurement

|  | NK | | SK | |
|---|---|---|---|---|
|  | Mean | SD | Mean | SD |
| Raven (Total 24) | 8.83 | 5.52 | 17.98 | 3.88 |
| Big5: Extraversion | 15.3 | 4.34 | 13.5 | 4.1 |
| Big5: Agreeableness | 13.94 | 3.65 | 13.65 | 3.47 |
| Big5: Openness | 15.15 | 3.6 | 14.35 | 3.6 |
| Big5: Conscientiousness | 12.23 | 3.72 | 14.74 | 3.88 |
| Big5: Neuroticism | 16.91 | 4.68 | 17.27 | 5.02 |
| Econ 101 | 3.54 | 1.70 | 5.20 | 1.33 |
| Financial Literacy | 1.41 | 0.87 | 2.17 | 0.78 |

# Online Appendix C.5. Trading behavior and Market profits

Here, we confirm our trading behavior is correlated with the double auction markets. As we can confirm in the Table C.5., We confirm that this market trading behavior is significantly correlated with their market profits at 5% level.

Table C.5. Trading Behavior and Market Profits

| VARIABLES | NK (1) Profits | SK (2) Profits |
|---|---|---|
| Targe Profits | 0.657*** | 0.754*** |
|  | (0.108) | (0.100) |
| Constant | 2.808 | 3.162** |
|  | (2.092) | (1.502) |
| Observations | 1,722 | 1,722 |
| R-squared | 0.647 | 0.710 |

Notes: In all specifications we control the group and period fixed effects. We also control gender, age, age squared, marital status. Human capital control includes standardized raven score, econ 101 test, financial literacy, Big 5 personality test.

We also report the same analysis for SK samples. We found that the target profit of SK sample is significantly correlated with actual profit in panel (a). However, unlike from NK sample, the correlation between IAT and target profit is not significant in most of markets.

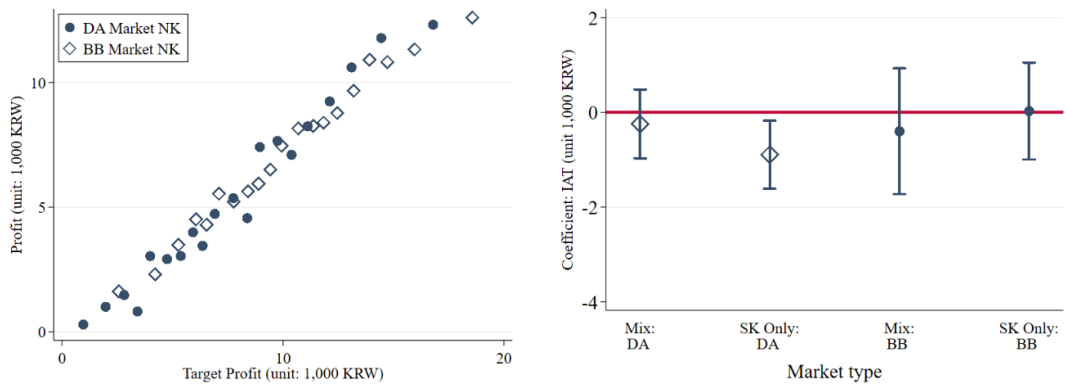

(a) Profit and Target Profit      (b) IAT and Target Profit

Fig C.5. Target Profit for SK

## Online Appendix C.6. Other Possible Mechanisms

In this section, we discuss other possible mechanisms for lower profits. We construct two different measures regarding active participation in the market: 1) Passivity and 2) a number of bids.

The first measure is about passive participation in the market. As a measure of a subject's passivity in trading, we simply define an indicator of whether the subject never submits any bid in a particular round. We first show that subjects who passively participate in the market have a lower level of profits. We note that we use the same set of controls as in the main text. In Table C.6.1., every specification is significant at 1% level which shows that passive participation is very important to explain the lower level of profits among NK refugees.

Table C.6.1. Passive participation and Market Profits

| Sample | NK (1) | NK (2) | NK (3) |
|---|---|---|---|
| | Dependent variable: Market Profits | | |
| Passive participation | -1.063*** (0.202) | -1.009*** (0.200) | -1.063*** (0.202) |
| Basic | N | Y | Y |
| Human Capital | N | N | Y |
| Experience in NK | N | N | Y |
| Adaptation in SK | N | N | Y |
| Observations | 1,722 | 1,722 | 1,722 |
| R-squared | 0.622 | 0.626 | 0.629 |

Note: The dependent variable is the amount of profit (1,000 KRW) in the DA market. Robust standard errors, clustered by individual, are presented in parentheses. */**/*** represent significance level 10/5/1\%. Group- and round-specific fixed effects are included in every specification. Basic controls include gender, age, age squared, and marital status. Preference controls include seller dummy, buyer's given value and seller's given cost, and risk preferences (average CE). Human capital control variables include standardized raven score, econ 101 test, financial literacy, and Big 5 personality measures. The variables of experiences in NK include lower economic class in NK, education level in NK, informal market experience, military service, communist party member, regions from bordering with China, and regions from Pyoungyang or Gaesung (big cities in NK). The variables about adaptation in SK include log income in SK, qualify of life in SK and whether they have family members left in NK.

We now investigate the effect of IAT on this passivity measure. In Fig C.6.1. most of the coefficients of IAT are positively correlated with passive participation which implies IAT could lower profits through passive market participation.

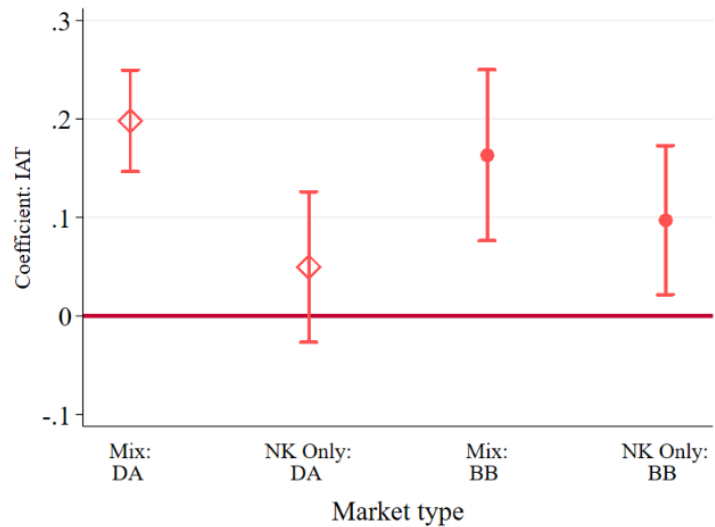

Fig C.6.1. IAT and Passive participation

The second measure is the number of bids and market profits. It is not clear that there is a linear relationship between the number of bids and market profits. For instance, subjects left behind with no deal can do a lot of bidding for the rest of the time. Therefore, we first run a regression using the number of bids and the squared number of bids as control variables.

Table C.6.2. Number of bidding and Market Profits

| Sample | NK (1) | NK (2) | NK (3) |
|---|---|---|---|
| | Dependent variable: Market Profits | | |
| # of bids | 0.457*** | 0.421*** | 0.451*** |
| | (0.122) | (0.119) | (0.119) |
| # of bids squared | -0.083*** | -0.079*** | -0.082*** |
| | (0.019) | (0.019) | (0.018) |
| Basic | N | Y | Y |
| Human Capital | N | N | Y |
| Experience in NK | N | N | Y |
| Adaptation in SK | N | N | Y |
| Observations | 1,722 | 1,722 | 1,722 |
| R-squared | 0.622 | 0.626 | 0.629 |

Note: The dependent variable is the amount of profit (1,000 KRW) in the DA market. Robust standard errors, clustered by individual, are presented in parentheses. */**/*** represent significance level 10/5/1\%. Group- and round-specific fixed effects are included in every specification. Basic controls include gender, age, age squared, and marital status. Preference controls include seller dummy, buyer's given value and seller's given cost, and risk preferences (average CE). Human capital control variables include standardized raven score, econ 101 test, financial literacy, and Big 5 personality measures. The variables of experiences in NK include lower economic

class in NK, education level in NK, informal market experience, military service, communist party member, regions from bordering with China, and regions from Pyoungyang or Gaesung (big cities in NK). The variables about adaptation in SK include log income in SK, qualify of life in SK and whether they have family members left in NK.

Using the coefficient in column (3), we find that there is a non-linear effect and the optimal number of bids is 2.8. We then construct a variable measuring the absolute gap from this optimal number of bids. Then we run a regression using IAT on this absolute gap from the optimal number of bids. We again find that as IAT increases, across markets the gap significantly increases in most of the markets in Fig C.6.2. This result shows that a number of bids could also be a possible mechanism that IAT lowers NK refugees' profit.

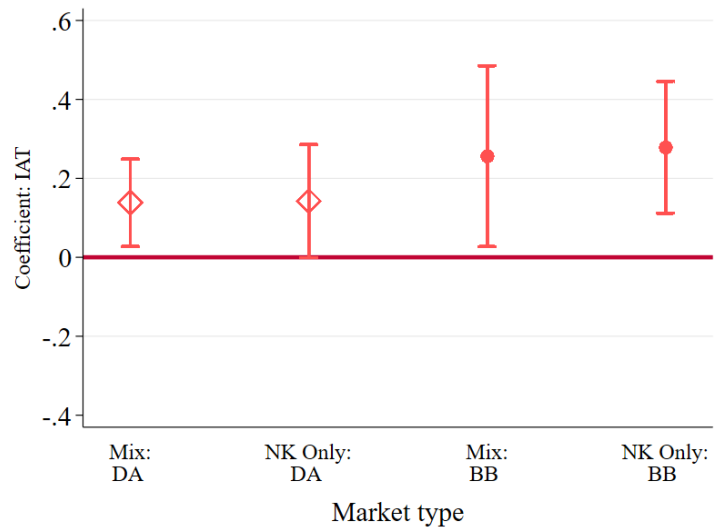

Fig C.6.2. IAT and number of bidding

## Online Appendix C.7. Robustness Check: Motivation

It is possible that the variation of market earnings in the experiment may reflect differences in motivation or preferences regarding effort provision in the artificial environment of the experiment. As a robustness check for our main results, we replicate our main Table 2 with two different additional controls: 1) the number of biddings and 2) passivity in the market. The passivity variable is a binary variable and 1 if subjects do not bid at all, while 0 if subjects bid at least more than once. We find that Table 2 is replicated in a robust way even after we control possible motivation issues in the experiment.

Table C.7.1. Controlling number of biddings in the market

| | A. Double Auction Market | | | |
|---|---|---|---|---|
| Market composition | All | Mix | NK only | SK only |
| | (1) | (2) | (3) | (4) |
| IAT | | | -0.707** | -0.338 |
| | | | (0.335) | (0.285) |
| IAT x NK | -0.586*** | -0.542** | | |
| | (0.194) | (0.236) | | |
| IAT x SK | -0.175 | -0.153 | | |
| | (0.193) | (0.241) | | |
| NK | -0.004 | -0.404 | | |
| | (0.220) | (0.282) | | |
| # of bidding | -0.095*** | -0.071* | -0.005 | -0.209*** |
| | (0.031) | (0.041) | (0.072) | (0.060) |
| Obs | 3,444 | 1,890 | 768 | 786 |
| R-sq | 0.648 | 0.664 | 0.611 | 0.688 |
| | B. Bilateral Bargaining | | | |
| | (5) | (6) | (7) | (8) |
| IAT | | | -0.887*** | 0.656 |
| | | | (0.341) | (0.468) |
| IAT x NK | -0.868*** | -1.150*** | | |
| | (0.290) | (0.422) | | |
| IAT x SK | 0.321 | -0.192 | | |
| | (0.414) | (0.668) | | |
| NK | -0.465 | -0.497 | | |
| | (0.458) | (0.808) | | |
| # of bidding | -0.425*** | -0.661*** | -0.350*** | -0.326*** |
| | (0.068) | (0.132) | (0.114) | (0.112) |
| Obs | 1,722 | 442 | 640 | 640 |
| R-sq | 0.324 | 0.388 | 0.368 | 0.288 |

Table C.7.2. Controlling passivity in the market

| | A. Double Auction Market | | | |
|---|---|---|---|---|
| Market composition | All | Mix | NK only | SK only |
| | (1) | (2) | (3) | (4) |
| IAT | | | -0.656** | -0.311 |
| | | | (0.331) | (0.289) |
| IAT x NK | -0.461** | -0.413* | | |
| | (0.189) | (0.230) | | |
| IAT x SK | -0.194 | -0.182 | | |
| | (0.192) | (0.237) | | |
| NK | 0.129 | -0.264 | | |
| | (0.215) | (0.270) | | |
| Passive participation | -0.646*** | -0.545*** | -1.315*** | -0.030 |
| | (0.147) | (0.195) | (0.308) | (0.310) |
| Obs | 3,444 | 1,890 | 768 | 786 |
| R-sq | 0.650 | 0.666 | 0.623 | 0.682 |
| | B. Bilateral Bargaining | | | |
| | (5) | (6) | (7) | (8) |
| IAT | | | -0.815** | 0.598 |
| | | | (0.344) | (0.456) |
| IAT x NK | -0.784*** | -1.006** | | |
| | (0.289) | (0.448) | | |
| IAT x SK | 0.224 | -0.425 | | |
| | (0.402) | (0.670) | | |
| NK | -0.118 | -0.046 | | |
| | (0.445) | (0.800) | | |
| Passive participation | 0.530 | 0.991 | 0.256 | 0.800 |
| | (0.374) | (1.022) | (0.449) | (0.787) |
| Obs | 1,722 | 442 | 640 | 640 |
| R-sq | 0.305 | 0.346 | 0.357 | 0.275 |

# Online Appendix D. Institutional Background

We introduce the institutional background about North and South Korea. Following the separation of the two Koreas, North Koreans have been subject to political education, instilling a socialist ideology. In contrast to capitalism based on private ownership and market coordination, socialism hinges on collective ownership of productive assets and coordination through central planning. These socialist premises of following instructions by central planners without claiming individual ownership require a mindset different from one in a market economy. The essence of such a socialist mindset is cooperation among economic agents. NK is not an exception, emphasizing collectivism and cooperation but perhaps with a stronger emphasis on these than any other socialist states (Oh and Hassig, 2004).

NK's official ideology is Juche, which is literally translated to "we-centeredness." Although the primary purpose of such an ideology is to enhance the pride of North Koreans as a unique socialist state and thus to justify a strong dictatorship ruled by one family, it also acts as a guiding principle in educating people and managing enterprises. "All for one and one for all" derived from the *Juche* ideology is a catch phrase that all North Koreans are expected to be aware of and behave in accordance with. Such an emphasis on collectivism is found in nearly all of the speeches of political leaders and in official documents. For example, Kim Il-sung, who is the founding father and the first political leader of NK, gave the following speech at a nursery school in 1966: "To become a communist, one should give up selfishness. … A communist must cherish the interests of a community and the society not her own interest." (Kim, 2001). In line with this ideology, an influential economics journal in NK attempts to elaborate the advantage of collectivism against capitalism with the following arguments (Ham, 1998): "If a man who is a member of community pursues money-making at a workplace with the motive of self-interest, not only the conflicts of interests among people take place but also unity and cooperation are destroyed." "In capitalism, men become vulgarians caring only for money and the society reduces to a jungle of competition where only the strongest survive." "A society in which people help each other and live as a member of a big family is … a socialist state."

To inculcate the socialist mindset in North Koreans, the NK authorities implemented a comprehensive system of ideological indoctrination. First, all North Koreans need to be members of official organizations whose objectives involve the teaching of the *Juche* ideology.

Children in the second year of elementary school and young people at the age of fourteen become members of the Boys' Union and Youth Union, respectively. Unions for adults are classified by occupations and all workers are affiliated with one of the unions. Adults without jobs are not exempt from a membership of an organization: housewives must belong to the Democratic Women's Union.

Second, all North Koreans are obliged to participate in weekly indoctrination meetings organized mainly by the above associations. During these meetings, participants are required to provide criticism of both themselves and of other participants using the supreme leaders' instructions as the absolute reference for correct attitude and behavior. In other words, such meetings aim to make North Koreans comply with the laws and regulations of the regime through both self-awareness and mutual surveillance.

Third, there are several other methods used to teach and discuss the instructions of political leaders and authorities. Families are encouraged to read such instructions at home, although that is often difficult to fully enforce. Workers are also required to read and discuss them before the start of work. In addition, many North Koreans visit places that function primarily as sites for the proliferation of propaganda. NK's labor law emphasizes the importance of learning the political ideology by stating "the principle of eight hours' work, eight hours' rest, and eight hours' learning should be fully enforced."

There is some evidence that North Koreans are affected by such programs of ideological indoctrination. According to the survey of NK refugees who newly arrived in SK in 2014, 54% replied that fellow North Koreans attended at least 70% of the indoctrination meetings (Jeong, 2015). Such rate of attendances might have been higher during the period before the 1990s, when NK began to experience a severe economic crisis. Furthermore, the same survey suggests that the average share of North Koreans who are proud of the *Juche* ideology amounted to 57% of total population. It is our key question in this study whether such intensive ideological indoctrination affects the mindset and the behavior of North Koreans.



\end{document}